\documentclass[10pt,a4paper,superscriptaddress,aps,prd,showkeys,showpacs,nofootinbib,reprint,preprintnumbers]{revtex4-1}
\usepackage{verbatim}
\usepackage{amsmath,amsfonts,amssymb}
\usepackage[breaklinks,colorlinks]{hyperref}
\usepackage{natbib}
\usepackage{tikz}
\usepackage[compat=1.1.0]{tikz-feynman}
\usepackage{float}
\usepackage{graphicx}
\usepackage{adjustbox}
\usepackage{tabularx}
\usepackage{tensor}
\usepackage{amsbsy}
\usepackage{braket}
\usepackage{slashed}
\usepackage{enumitem}
\hypersetup{%
,urlcolor=blue
,citecolor=blue
,linkcolor=blue
}

\begin{document}
\title{ { \large Superradiance in Stars: }\\
\normalfont{Non-equilibrium approach to damping of fields in stellar media}
}

\author{F.~Chadha-Day}
\email[]{francesca.chadha-day@durham.ac.uk}
\affiliation{Institute for Particle Physics Phenomenology, Department of Physics, Durham University,
Durham DH1 3LE, United Kingdom}

\author{B.~Garbrecht}
\email[]{garbrecht@tum.de}
\affiliation{%
Technische Universität München, Physik-Department, James-Franck-Straße, 85748 Garching, Germany
}

\author{J.~I.~McDonald}
\email[]{jamie.mcdonald@uclouvain.be}
\affiliation{Centre for Cosmology, Particle Physics and Phenomenology,
Université Catholique de Louvain,
Chemin du cyclotron 2,
Louvain-la-Neuve B-1348, Belgium}
\affiliation{Institute for Particle Physics Phenomenology, Department of Physics, Durham University,
Durham DH1 3LE, United Kingdom}
\preprint{IPPP/22/46}

\label{firstpage}

\date{\today}

\begin{abstract}
%Superradiance in black holes is well-understood but a general treatment for superradiance in stars has until now been lacking. This is surprising given the ease with which we can observe isolated neutron stars and the array of signatures which would result from stellar superradiance. In this work, we present the first systematic pipeline for computing superradiance rates in rotating stars starting from any fundamental Lagrangian describing the interaction between the superradiant field and the constituents of the star. Our scheme falls into two parts:  firstly we show how field theory at finite density can be used to express the absorption of long wavelengths into the star in terms of microphsyical scattering processes. This allows us to derive a damped equation of motion for the bosonic field. We then feed this into an effective theory for long wavelengths (the so-called worldline formalism) to describe the  amplification of superradiant modes of arbitrary multipole mode for a rapidly rotating star.

Superradiance in black holes is well-understood but a general treatment for superradiance in stars has until now been lacking. This is surprising given the ease with which we can observe isolated neutron stars and the array of signatures which would result from stellar superradiance. In this work, we present the first systematic pipeline for computing superradiance rates in rotating stars.
Our method can be used with any Lagrangian describing the interaction between the superradiant field and the constituents of the star. Our scheme falls into two parts:  firstly we show how field theory at finite density can be used to express the absorption of long wavelength modes into the star in terms of microphsyical scattering processes. This allows us to derive a damped equation of motion for the bosonic field. We then feed this into an effective theory for long wavelengths (the so-called worldline formalism) to describe the amplification of superradiant modes of arbitrary multipole moment for a rapidly rotating star. Our method places stellar superradiance on a firm theoretical footing and allows the calculation of the superradiance rate arising from any interaction between a bosonic field and stellar matter.

\end{abstract}



\maketitle

\section{Introduction}

% Trento talk Tuesday Main hall . Andreas Schmidtt, Geraint. Southampton Tuesday Main hall. Interior structure of neutron star.
As we search for clues of new physics, astroparticle probes have in recent years experienced an explosion of interest. It is becoming increasingly apparent that physics beyond the standard model may lead to a vast array of new and spectacular signatures arising from stars, black holes and galaxies \cite{Berti:2022rwn}.  This shift is driven both by the non-observation of new physics at colliders (especially dark matter) and the advent of multi-messenger astronomy, where the last decade has seen the emergence of gravitational wave detectors and increasingly sophisticated X-ray and radio telescopes.  

In this regard superradiance \cite{Brito:2015oca,Bekenstein:1998nt} sits perfectly at the intersection of many of these fields. Superradiance is a general phenomenon which manifests itself in many systems. In the present context, however, we are concerned with so-called \textit{rotational superradiance}. This variety of superradiance concerns the extraction of rotational energy from fields scattering off spinning objects. This occurs when low frequency waves are re-scattered with an amplitude larger than an incident wave providing there is some mechanism for energy exchange with object. This was originally illustrated by Zel’Dovich \cite{zeldovich1,zeldovich2} for the case of electromagnetic waves scattering off a rotating body with finite conductivity (losses). This same mechanism manifests itself for rotating Kerr black holes, where energy exchange is provided by dynamics occurring close to the black hole\footnote{There is some discussion about whether or not superradiant scattering requires a horizon, or simply the presence of an ergo region \cite{Brito:2015oca,Vicente:2018mxl,Eskin:2015ssa,Richartz:2009mi}. For our purposes this is however a moot point, since the ergoregion would be inside the star, where in any case one should use an interior metric different from pure Kerr. We therefore investigate other sources of direct dissipation coming from collisions of the matter in the stellar interior.}.  Compact objects offer one additional ingredient, namely that massive fields can become trapped around the black hole owing to the existence of gravitational bound state solutions in black hole geometries. This trapping and continual extraction of rotational energy leads to unstable bound states for massive fields whose amplitude grows exponentially in time. This is black hole superradiance. Given that superradiance will generically occur for any rapidly rotating compact object, an immediate question is whether the same mechanism will occur also in rapidly rotating neutron stars.

As with black holes, superradiance in stars would lead to a variety of striking phenomena including gravitational waves \cite{Arvanitaki:2014wva,Brito:2017wnc,Baryakhtar:2017ngi,Brito:2017zvb}, polarisation signatures \cite{Chen:2019fsq} and effects on binary inspirals \cite{Baumann:2018vus}. Superradiance can also be used to constrain the existence of light fields \cite{Arvanitaki:2010sy,Pani:2012vp,Cardoso:2018tly}. It is also worth remarking on a recent observation that the standard model photon itself may experience superradiance growth around black holes owing to its finite plasma mass \cite{Blas:2020kaa,Wang:2022hra,Conlon:2017hhi}

Superradiance in stars has been studied for some specific cases \cite{Cardoso:2015zqa,Cardoso:2017kgn,Day:2019bbh} but a general treatment akin to that for black holes has until now not been formulated. Here we present a framework for computing the superradiance rates in stars given any fundamental interaction.

Our key insight is to employ techniques from quantum field theory at finite temperatures and densities. This allows us to describe, in a very general way, an effective equation of motion for the bosonic field. This equation describes the interaction between long-wavelengths of the superradiant field and the constituents of the star, providing the missing link between microphysics and damping terms that arise in the equations of motion for the boson.

This result is a satisfying illustration of the power of formal techniques in field theory to answer questions in astrophysics. Such tools are more usually brought to bear in the  Early Universe to describe systems at finite temperature and out-of-equilibrium (most usually in homogeneous but time-varying environments). Applying these same procedures in an astrophysical setting is quite novel. Here one must come to terms with the fact that the star may provide a stationary, but not spatially invariant background.

Once an effective equation of motion is derived, the superradiance rate then follows in a straightforward way by solving this equation for spherical modes and mapping this onto superradiant bound states. Below we summarize our scheme in more detail.

% SIGNATURES/CONSTRAINTS IF IT WORKS.  

\section{Summary}

The rate of stellar superradiance depends crucially on the {\it damping} rate of the potentially superradiant field $\phi$ in the stellar background. In Section~\ref{sec:NeqQFT}, we show how nonequilibrium field theory allows us to derive an effective damped equation of motion for the scalar field
   \begin{equation}\label{eq:phiEQ}
       \partial^2 \phi  + \mu^2 \phi + \Gamma_\phi \dot{\phi} = 0,
   \end{equation}
where the damping is given by $\Gamma_\phi =- \lim_{p\rightarrow 0} \text{Im} \, \Pi^R(p)/p^0$ evaluated in the long-wavelength limit $p\rightarrow 0$, where $\Pi^R$ is $\phi$'s in-medium self-energy
  \begin{center}
    \includegraphics[scale=0.76]{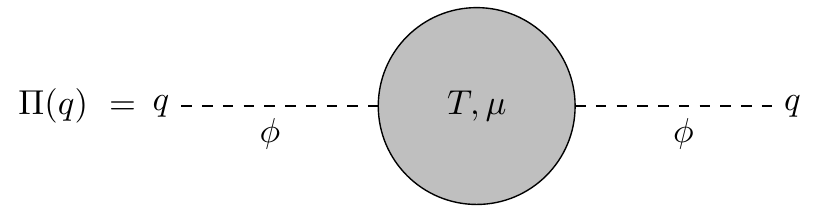}.
    \end{center}
Despite the simple appearance of Eq.~\eqref{eq:phiEQ}, it is worth emphasising that demonstrating the validity of this result from first principles in non-equilibrium quantum field theory and relating the damping rate to Green functions and thence to collision integrals is a non-trivial and powerful result, relying on a series of controlled approximations which are explained step-by-step in later sections.  This is testament to the power of fundamental QFT and represents a novel application of finite density field theory to a problem in astroparticle physics.

Below we summarise our scheme for computing superradiance rates in stars, which consists of the following steps:

\begin{itemize}[leftmargin=*]

\item The first task is to calculate the imaginary part of the self energy that yields $\Gamma_\phi$ per the above relation. This can be computed directly within the framework of thermal field theory or, under approximations discussed below, related via the optical theorem to a collision integral

    \begin{align}\label{eq:CollisionIntegral}
      &\text{Im} \, \Pi(q) =\nonumber \\
      &\prod_i \int \frac{d^3\textbf{p}_i}{(2\pi)^3 2E_i} f_i \prod_j \int \frac{d^3\textbf{p}_j}{(2\pi)^3 2E_j} (1 \pm f_j) \left|\mathcal{M}\right|^2,
    \end{align}
    \begin{center}
    \includegraphics[scale=0.7]{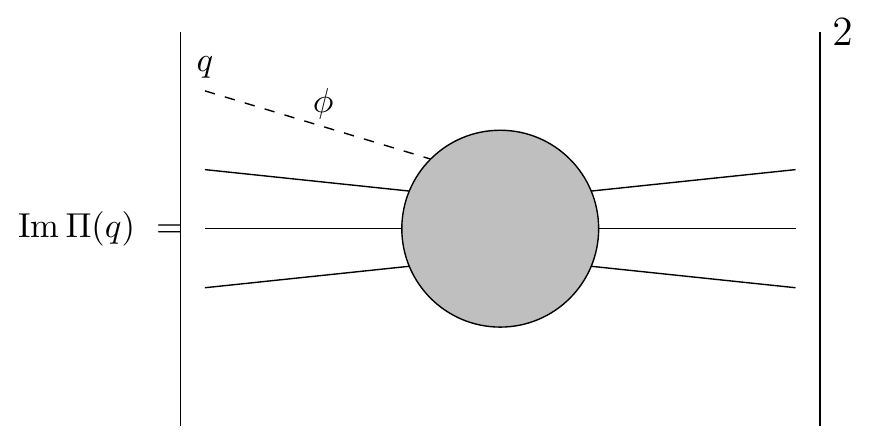}
    \end{center}

where $f_i$ are distribution functions for particles in the medium and $\mathcal{M}$ is a scattering amplitude. 
% \item We are now in a position to write down the equation of motion for $\phi$ inside the star in the long wavelength limit,
%   \begin{equation}\label{eq:phiEQ}
%       \partial^2 \phi  + m_a^2 \phi + \Gamma_\phi(0) \dot{\phi} = 0.
%   \end{equation}
% where $\Gamma_\phi(p) = \text{Im} \, \Pi^R(p, x)/p^0$ evaluated in the long-wavelength limit $p\rightarrow 0$ in Eq.~\eqref{eq:phiEQ}.
    % {\color{red} Is it true that $\text{Im} \Pi^R = \text{Im} \, \Pi(q)$? If so, can we stick with $\text{Im} \, \Pi(q)$ in the summary to avoid lots of notation?}
    
    From the classical equation of motion \eqref{eq:phiEQ}, one can derive an expression for scattering of an arbitrary mode $(\ell, m , \omega)$ into a non-rotating star of radius $R$
    
    \begin{equation}
Z_{\ell m } \simeq -\frac{4 R^2   \omega (R \omega)^{2\ell + 1}}{(1+ 2\ell )!! (2\ell + 3 )!!}\cdot\Gamma_\phi.
\label{eq:ZlmStarSummary}
\end{equation}

\item The superradiant instability rate of a massive field gravitationally bound to the star can then be calculated in the world line effective field theory framework \cite{Endlich:2016jgc}, as described in Section \ref{sec:EFT}. The scattering described by equation \eqref{eq:ZlmStarSummary} is essential in this process to match the effective field theory to the low energy physics of the star. The superradiant growth rate for harmonic  bound state labeled by $(n,\ell,m)$
\begin{align}\label{eq:SupRatePi}
    &\Gamma_{n \ell m} = C_{n \ell m} \left(\frac{R}{r_{n \ell}} \right)^{(2\ell +3)}  \frac{(m \Omega- \omega_{\ell n} )}{\omega_{\ell n}} \Gamma_\phi ,
    %   &\nonumber  {\normaltext \text{\quad \textcolor{purple}{Fran check}} }
\end{align}
where $C_{n \ell m}$ is a combinatorial coefficient defined later on in Eq.~\eqref{eq:Cnlm}, $r_{n\ell} = (n+\ell+1)/(2GM \mu^2)$ gives the characteristic radius of the superradiant profile, and $\omega_{\ell n}$ gives the real part of the discrete bound state frequencies Eq.~\eqref{eq:BSFreqs}. The rotational frequency of the star is $\Omega$. 

\end{itemize}

\section{Dissipation of Fields in QFT}\label{sec:NeqQFT}

Dissipation appears ubiquitously and is of importance for many physical systems. This is also true for particle physics. While the phenomenon is universal, there are many different calculational approaches, and the choice of these can have various motivations. The following methods are of relevance in the present discussion:
\begin{itemize}
    \item Given the utility of scattering theory in particle physics, inferring the damping rate of a field from cross sections and the density of the medium is a straightforward way that is often pursued. Limitations are met when other medium effects such as corrections to the dispersion relations or screening are of importance.
    \item The Matsubara formalism for finite-temperature (equilibrium) field theory can be applied to compute Green's functions from which the damping rate and other medium effects can be inferred. In the present context, it is of interest whether such linear response coefficients also apply to the wave equation for a classical field $\phi$.
    \item The Schwinger-Keldysh closed-time-path (CTP) approach  \cite{Keldysh:1964ud,Schwinger:1960qe} allows one in principle to calculate the real time evolution of the classical field and of correlation functions. In practice however, far-reaching approximations and truncations must be applied to the equations that directly descend from first principles. Such approximations result in descriptions which can be solved numerically or analytically.
\end{itemize}

In the present work, we carry out some developments on the CTP approach in order to apply these to the damping of scalar field waves. In Subsection~\ref{subsec:CTP}, we review the CTP formalism, where the purpose is to declare the basic conventions as well as to provide a very basic introduction to the matter. Reviews and research articles that offer elaborations that are both, more detailed and more synoptic, include Refs.~\cite{Berges:2004yj,Calzetta:2008iqa,Calzetta:1986cq,Prokopec:2003pj}. We then proceed to derive equations of motion for a inhomogeneous scalar field $\phi$ in a general background to linear approximation in $\phi$. These can be viewed as modified wave equations that include dispersive and absorptive effects. Complementary to this are studies where the evolution of a spatially homogeneous condensate is treated to nonlinear level also using CTP methods~\cite{Ai:2021gtg,Cheung:2015iqa}

In Subsection~\ref{subsec:Wigner}, we carry out the transformation to Wigner space, which allows to truncate the corrections to the wave equations so that these are local, i.e. they do not feature memory integrals. In other words, the integro-differential equations are approximated by the a wave equation in the usual form of the partial differential equation~(\ref{eq:PiDiamond}). We discuss various additional approximations, in particular the truncation of the gradient expansion, as well.

\subsection{Closed-time-path approach and equations of motion}\label{subsec:CTP}

When first introduced to the subject of quantum field theory one begins by considering fluctuations about the quantum vacuum, where physical observables are associated to the expectation values of some set of operators. The basic objects in any field theory are the two-point functions, for instance, the Wightman function defined by
\begin{equation}\label{eq:in-out}
    \braket{\varphi(x)  \varphi(y) }_{\rm vac.} = \bra{0,t \rightarrow \infty} \varphi(x)  \varphi(y) \ket{0, t\rightarrow -\infty}
\end{equation}
where $\ket{0}$ is the ground state of the theory. Further, one assumes that the one-point function of a generic field $\varphi$ will have a constant expectation value $\phi(x)=\braket{\varphi(x)} =\text{const.}$, there are no statistical fluctuations and the system is entirely described by a pure vacuum state. Generally $\varphi$ may refer here to an arbitrary number of fields, so that there may be flavour and Lorentz tensor or spinor structure that we suppress for the time being.

By contrast, many physical systems are in a state that is quite different from the vacuum $|0\rangle$. Of interest can be situations with a finite density of particles,  described in general by a mixed (possibly thermal) state. Fields may attain non-trivial, spacetime-dependent, expectation values $\phi=\braket{\varphi} \neq \text{const.}$ which vary throughout time and space. The evolution of these expectation values can influence and be influenced by the presence of other degrees of freedom. The latter may or may not be close to thermal equilibrium. Clearly such systems require some self-consistent treatment to address the interconnected dynamics of each field and its out-of-equilibrium evolution.

\begin{figure}
    \centering
    \includegraphics[scale=0.75]{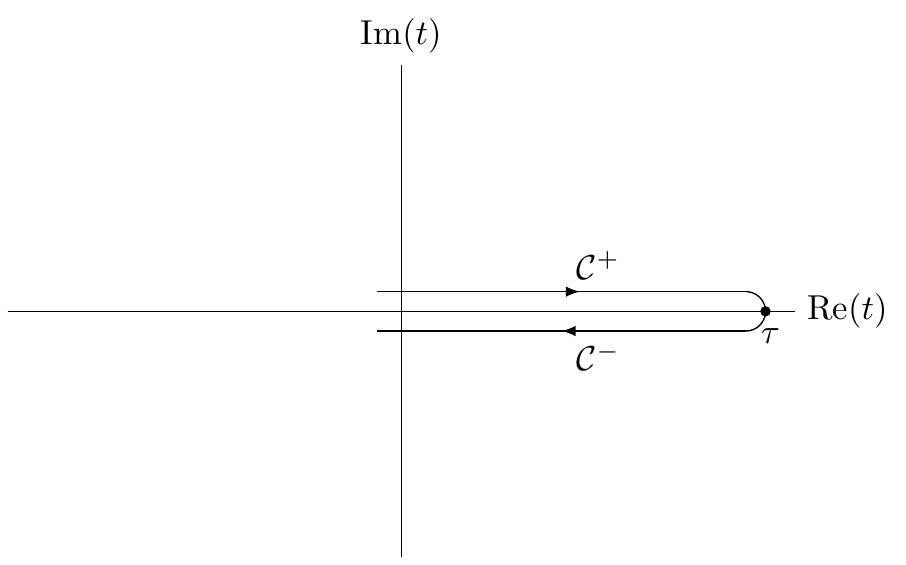}
    \caption{The Schwinger Keldysh Contour. The closed path in the complex time domain along which fields are evolved, corresponding to Eq.~\eqref{Z:in:in}, where the time $t$ corresponds to $x^0$.}
    \label{fig:CTP}
\end{figure}

The description of such systems rests on two fundamental points: that they evolve unitarily according to a Hamiltonian $\mathcal{H}$ and that the state of the system is described by a density matrix $\rho$. (Often, unitarity is not respected by truncations. If the resulting equations describe e.g. the state of a thermodynamic system they are typically irreversible.) In such systems, generalized correlators are defined by
\begin{equation}
    \braket{\varphi(x) \varphi(y) }  = {\rm tr} \left[ \rho(t) \varphi(x) \varphi(y)  \right]
\end{equation}
with similar definitions existing for the other Wightman function (with exchanged coordinates), and the time-ordered (Feynman) and anti-time-ordered (anti-Feynman) correlation functions.  Such correlations ought to be expressible via functional methods by taking derivatives of an appropriately defined generating functional. Clearly the trace necessitates matrix elements where the ket and bra states are the same - the so-called ``in-in" scenario. This is in contrast to Eq.~\eqref{eq:in-out} where the ingoing and outgoing states are defined at different times. These are instead referred to as ``in-out" expectation values. Furthermore, we know that matrix elements can be re-expressed in terms of path integrals. Matrix elements of the ``in-in" type, where both ket and bra states are evaluated at the same time $t$, can be evaluated by inserting a complete basis of states at some intermediate time $\tau$. The path integral can then be used to evolve forward from $t$ to $\tau$ and then backwards again in the reverse direction~\cite{Schwinger:1960qe,Keldysh:1964ud}, forming a closed loop in the time-domain, as illustrated in FIG.~\ref{fig:CTP}. This corresponds to the following generating functional~\cite{Calzetta:1986cq,Calzetta:2008iqa}
\begin{align}
&Z[J_+,J_-]\notag\\=&\int\cal{D \varphi(\tau)}{\cal D}\varphi^-_{\rm in}{\cal D}\varphi_{\rm in}^+\langle \varphi^-_{\rm in}|\varphi(\tau)\rangle
\langle\varphi(\tau)|\varphi_{\rm in}^+\rangle \langle\varphi^-_{\rm in}|\varrho|\varphi_{\rm in}^+\rangle
\notag\\
=&\int{\cal D}\varphi^-{\cal D}\varphi^+
{\rm e}^{{\rm i}\int d^4 x\{{\cal L}(\varphi^+)-{\cal L}(\varphi^-)+J_+\varphi^+-J_-\varphi^-\}}\langle\varphi^-_{\rm in}|\varrho|\varphi_{\rm in}^+\rangle
\,.
\label{Z:in:in}
\end{align}
It generates all correlators, in particular the one-point function
\begin{equation}
    \phi = \braket{\varphi} = \pm \left. \frac{\delta Z[J_+,J_-]}{\delta J_\pm} \right|_{ J_{\pm} = 0}
\end{equation}
but also the set of four Green functions
% (the bases are two dimensional and are given by e.g. one of the Wightman functions and one (anti-)time ordered correlator)
\begin{align}\label{eq:Correlators}
{\rm i}\Delta^{ab}(x,y)=&-\frac{\delta^2}{\delta J_a(x) \delta J_b(y)}\log Z[J_+,J_-]\Big|_{J_\pm=0}\notag\\
=&{\rm i}\langle{\cal C}[\varphi^a(x)\varphi^b(y)]\rangle\,,
\end{align}
where we refer to $a,b = \pm$ as Schwiger--Keldysh indices and $\mathcal{C}$ defines a path ordering along the so-called Schwinger Keldysh contour in FIG.~\ref{fig:CTP}. We can then generate
\begin{align}
\label{CTP:indices}
G^>(x,y)=&G^{-+}(x,y)\,,\quad G^<(x,y)=G^{+-}(x,y)\,,\notag\\
G^T(x,y)=&G^{++}(x,y)\,,\quad G^{\bar T}(x,y)=G^{--}(x,y)\,,
\end{align}
which are the two Wightman and time and anti-time ordered propagators, respectively. In turn, linear combinations of these give the retarded and advanced Green functions
\begin{align}
G^{\rm r}=&G^T-G^<=G^>-G^T\,,\notag\\\quad G^{\rm a}=&G^T-G^>=G^<-G^{\bar T}. \label{eq:RetardedAdvanced}
\end{align}
We write here $G$ instead of $\Delta$ for generality because these definitions apply as well for other two-point functions than correlators, in particular for self energies.

% \begin{equation}
%      \braket{\mathcal{O}(t)}   = \bra{\alpha} \rho(t) \mathcal{O}  \right]
% \end{equation}
% For systems in thermal equilibrium or in vacuum, observables are given simple by their expectation values owing to thermal or quantum fluctuations about their  

%\begin{align}
%G^{\cal A}=\frac{\rm i}{2}(G^>-G^<)=\frac{1}{2{\rm i}}(G^{\rm a}-G^{\rm r})
%\end{align}
%
%\begin{align}
%G^{\rm H}=\frac{1}{2}(G^{\rm a}+G^{\rm r})\,.
%\end{align}
%
%\begin{align}
%\label{T:T Bar:gr:le}
%G^T+G^{\bar T}=G^>+G^<\,.
%\end{align}

% ADD ALSO 1PI HERE ....

Although we have presented a formalism for computing correlation functions for fields out of equilibrium, we need to go one step further in actually describing their real-time evolution through effective equations of motion. This is especially true of the one-point function $\phi=\braket{\varphi}$, whose equation of motion will eventually be identified with an effective description for the dynamics of a superradiant field. What we need then is a toolkit for describing the interconnected evolution of correlators. For this we shall use the two-particle irreducible (2PI) effective action formalism, which allows us to produce an intercoupled set of equations for one and two-point functions.

The 2PI effective action \cite{Cornwall:1974vz,Calzetta:1986cq,Berges:2004yj} has been established as a powerful formalism for providing a description of the real time evolution of nonequilibrium systems, where the relevant dynamics is typically described by two-point functions. For the present work, the main target is the evolution of the one-point function $\phi$, and backreaction to the finite density system is neglected. For this reason, the relevant equations of motion could also be derived here from the generating functional or the one-particle effective action. However, this would not amount to a substantial simplification, and in view of applications to nonequilibrium systems accounting for backreactions from $\phi$, we carry out the formal developments here in the 2PI framework.
% Typically, e.g. for plasmas at many different temperatures and chemical potentials, observables that can be derived from two-point correlators are of primary interest. In the present work, the main target is the evolution of the one-point function $\phi$, which in the context of effective actions is often referred to as the classical field.
% The 2PI formalising describes a powerful formalism for describing the self-consistent dynamics of a field and its response function, as embodied by the 2-point function. For reviews see \cite{Cornwall:1974vz,Berges:2004yj}.

The 2PI effective action for a number of fields $\varphi$  and their propagators $\Delta$ (where flavour, tensor, spinor and Schwinger--Keldysh indices are suppressed) is given by
% \begin{align}
% \label{Gamma2PI}
% \Gamma_{2\rm PI}[\phi,\Delta]=S+{\rm i}\,{\rm tr}[{\Delta^{(0)}}^{-1}\Delta]
% +{\rm i}\,{\rm tr}\log\Delta^{-1}+\Gamma_2[\phi,\Delta]
% \end{align}
\begin{align}
\label{Gamma2PI}
&\Gamma_{2\rm PI}[\phi,\Delta] \nonumber \\
=&S[\phi]+  {\rm i}\,{\rm tr}[{\Delta^{(0)}}^{-1}\Delta]
+{\rm i}\,{\rm tr}\log\Delta^{-1} +\Gamma_2[\phi,\Delta].
\end{align}
where $S$ is the classical action and  ${\rm i}{\Delta^{(0)}}^{-1}=\delta S/(\delta\phi\delta\phi)$ is the classical inverse propagator and $\Gamma_2$  is given by $(-{\rm i})$ times the sum of the 2PI vacuum graphs.
% \begin{equation}
%   {\Delta^{(0)}}^{-1} = \partial^2 - V_{,\phi \phi}(\phi) +
% % \end{equation}
% \begin{align}
% \label{Gamma2}
% [\phi,\Delta]\equiv -{\rm i}\times \textnormal{the sum of 2PI vacuum graphs}\,,
% \end{align}
The traces are then to be taken over all indices, whether they pertain to flavour, tensor, spinor structures or to the Schwinger--Keldysh contour. Note that generally, neither ${\rm i}\Delta^{(0)}$ nor ${\rm i}\Delta$ are diagonal in flavour but, of course, spacetime symmetry prohibits the mixing of fields from different representations of the Lorentz algebra. Finally, the traces are understood to act in complex field space, so that for a real scalar field, a factor of $1/2$ is implied.

A closed set of equations of motion follows from the stationary points in the functional dependence of $\Gamma_{2\rm PI}$ on $\phi$ and $\Delta$:
\begin{align}\label{eq:stationary}
&\frac{\delta\Gamma_{2\rm PI}[\phi,\Delta]}{\delta\phi(x)} = 0,
  \qquad \frac{\delta\Gamma_{2\rm PI}[\phi,\Delta]}{\delta \Delta(x,y)}=0 .
\end{align}
Again, all indices are suppressed here. Note that for the second of these equations, there are four different Schwinger--Keldysh tensor components. These can be decomposed into the celebrated Kadanoff--Baym equations as well as equations for the retarded or advanced Green functions, see e.g. Ref.~\cite{Prokopec:2003pj}.
 
 In order to illustrate how dissipation of a classical field can result from its interactions with another sector, we consider a simple model consisting of a bosonic field  $\varphi$ coupling to a massive fermion $\psi$ via
% \begin{equation}
%     \mathcal{L} = \frac{1}{2} \partial^2 \varphi +V(\varphi) + i \bar{\psi} \slashed{\partial}\psi + m_\psi \bar{\psi} \psi + \lambda \varphi \bar{\psi} e^{i\alpha\gamma^5} \psi.
% \end{equation}
\begin{equation}
    \mathcal{L} = \frac{1}{2} \partial^2 \varphi +V(\varphi) + i \bar{\psi} \slashed{\partial}\psi + m_\psi \bar{\psi} \psi + \lambda \varphi \bar{\psi}  \psi.
\end{equation}
(To roll back to the above notation, refer to the scalar and the fermion field collectively as $\varphi$, i.e. $\left\{\varphi,\psi\right\}\to \varphi$.)
This provides a toy proxy for the kind of interactions between a (pseudo)scalar field and the constituents of the star which we will consider in subsequent sections. It should also serve as an example to clarify how to derive equations of motion for scalar fields subject to different interactions.

We are principally interested in how the scalar dynamics is affected by the presence of the second, fermionic field. This is described by the first equation in \eqref{eq:stationary}, i.e. $\delta\Gamma/\delta\phi =0$, leading to
\begin{align}
&\partial^2\phi(x)+V_{,\phi}(\phi(x))  \nonumber \\ &+\frac{\delta\left(i{\rm tr}[{\Delta^{(0)}}^{-1}\Delta]+
i{\rm tr}[{{\cal S}^{(0)}}^{-1}{\cal S}]+\Gamma_{2}\right)}{\delta \phi} =0,\label{eq:EOM:exact:phi}
% \frac{\delta\Gamma[\phi,\Delta]}{\delta\Delta(y,x)}=&0
% \Leftrightarrow
% &{\rm i}{\Delta^{(0)}}^{-1}(x,y)
% -{\rm i}\Delta^{-1}(x,y)-{\rm i}\Pi(x,y)=0 .
\end{align}
where
\begin{align}
i\Delta^{(0)\, -1} &= \partial^2 + V_{,\phi\phi}(\phi),  \label{eq:Delta0Inverse}\\
i {\cal S}^{(0) \, -1} &= i\slashed{\partial}  + m_\psi + \lambda \phi
% e^{i\alpha \gamma^5}
\label{eq:S0Inverse}
\end{align}
and ${\cal S}$ and $\Delta$ are the full fermionic and bosonic propagators, respectively. The term with the derivative $\delta/\delta\phi$ in Eq.~\eqref{eq:EOM:exact:phi} introduces the effect of loop corrections on the dynamics of $\phi$. We take the potential to be of the form
\begin{equation}
    V(\phi) = \frac{\mu^2}{2} \phi^2.
\end{equation}
From Eq.~\eqref{eq:Delta0Inverse} we see then that $\Delta^{(0)\, -1}$ has no explicit $\phi$ dependence and is therefore annihilated by the action of the functional derivative with respect to $\phi$. (Recall that the fields $\phi$ and the propagators $\Delta$ are independent variables in the 2PI effective action.) Furthermore, for the Yukawa interaction, insertions of $\phi$ into diagrams are absorbed within the fermion propagator, cf. Eq.~(\ref{eq:S0Inverse}). Hence $\Gamma_2$ does not explicitly depend on $\phi$ and does not contribute to the derivative term in Eq.~(\ref{eq:EOM:exact:phi}). (Generally, when there are explicit insertions of $\phi$ into vertices of 2PI diagrams, the terms from $\Gamma_2$ enter the equations of motion as well.) Therefore, in the present example, only the term ${\rm tr}[{{\cal S}^{(0)}}^{-1}{\cal S}]$ survives the functional differentiation with respect to $\phi$, which appears in ${{\cal S}^{(0)}}^{-1}$, leaving us with
\begin{align}
&\partial^2\phi(x)+ \mu^2 \phi(x) + \lambda
% e^{i\alpha\gamma^5}\,
i {\rm tr}[{\cal S}(x,x)] =0\,.
\end{align}

Now, while the fermion propagator $i {\cal S}$ does not explicitly depend on $\phi$, it does so implicitly through its equation of motion in which $\phi$ appears. When $i {\rm tr}[{\cal S}(x,x)]_{\phi=0}\not=0$ (i.e. for constant, vanishing field configurations), the fermion loop shifts the time-independent solution for the expectation value of $\phi$. This contribution may be removed by adding appropriate counter terms or simply by a redefinition of the field $\phi$ by a constant value.%Another option is to introduce a global Peccei--Quinn symmetry that protects the potential of $\phi$ against radiative corrections from fermions.

\begin{figure}
    \centering
    \includegraphics[scale=0.8]{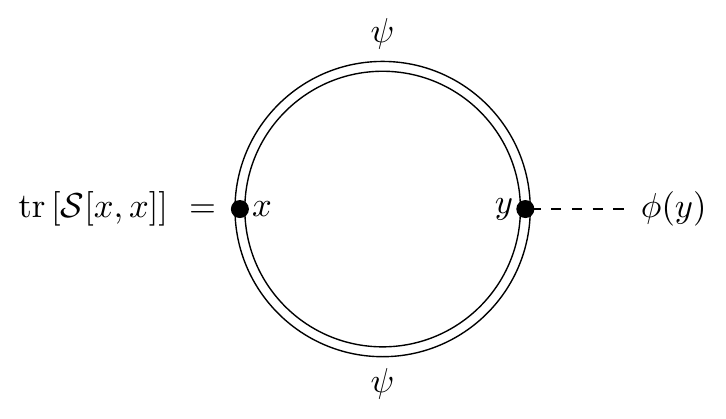}
    \caption{Tadpole diagram contributing to ${\rm tr}[S(x,x) ]$ at linear order in $\phi$. Double lines denote full fermion propagators in the event that $\psi$ is coupled to some other sector different from $\phi$. Similarly, thick dots may also represent vertex corrections from such a sector.}
    \label{fig:tadpole}
\end{figure}

Assuming that the dynamics is adequately described by a linear approximation,\footnote{At later stages of evolution, when field amplitudes become large, non-linearities can enter, and bosenova collapse can occur for superradiant fields \cite{Yoshino:2012kn,Yoshino:2013ofa} in $\phi$, but for now we are only interested in addressing superradiance with a linear stability analysis.} we functionally expand the implicit dependence of the solution for $i {\cal S}$ to linear order in $\phi$.  That is, we approximate
\begin{equation}
     \lambda\,
    %  e^{i\alpha\gamma^5}\,
{\rm tr}[{\cal S}(x,x) ] = \int d^4y \Pi(x,y) \phi(y),
\end{equation}
where $\Pi$ is simply the scalar self energy evaluated in the background $\phi(x)\equiv 0$. The resulting term in the wave equation is represented by the Feynman diagram in Fig.~ \ref{fig:tadpole}. This leads to the following form of the wave equation:
% \begin{equation}
%         \partial^2 \phi^a(x) + \mu^2 \phi^a(x) +
%     \int dy \, b \Pi^{ab}(x,y) \phi^b(y) = 0.
% \end{equation}
\begin{equation}
        \partial^2 \phi^a(x) + \mu^2 \phi^a(x) +
    \int d^4y \,  \Pi^{ab}(x,y) h^{bc} \phi^c(y) = 0.
\end{equation}
where we have now made the Schwinger--Keldysh indices $a,b,c=\pm$ explicit, with $h^{ab}$ defining a Schwinger-Keldysh metric given by  $h={\rm diag}(1 ,-1)$ which enforces the signs inherent in Eqs.~\eqref{Z:in:in}--\eqref{eq:Correlators}. Taking $a=+$, one can see that by  using $\phi^+ =\phi^- = \phi$ and the definitions  \eqref{CTP:indices} and \eqref{eq:RetardedAdvanced} generalized to self-energies, the final term can be expressed as
\begin{equation}\label{eq:ScalarEOM}
        \partial^2 \phi(x) + \mu^2 \phi(x) +
    \int d^4 y \, \Pi^R(x,y) \phi(y) = 0,
\end{equation}
where $\Pi^R(x,y) = \Pi^{++} - \Pi^{+-}$ is the retarded self-energy. Equations of this form are similar to those derived in Refs.~\cite{Cheung:2015iqa,Ai:2021gtg} for homogeneous field configurations. The fact that a memory integral, i.e. a convolution with the retarded self energy, appears here implies that this term in principle depends on values of $\phi(y)$ with $y^0<x^0$. It may therefore be characterized as non-Markovian.

The fact that the convolution of $\phi$ with the retarded self energy appears in Eq.~(\ref{eq:ScalarEOM}) is a generic feature when the loop propagators lineraly depend on $\phi$. Another important situation is the quadratic dependence, i.e. for a quartic self interaction $\varphi^4$ or a coupling to another scalar $\chi$ via $\varphi^2\chi^2$. Then, tadpole corrections of the form
\begin{align}
\Pi^{\rm TP}(x)\phi(x)\nonumber
\end{align}
appear in the equations of motion. With the help of a $\delta$-function, these can be of course brought to the form of the self energy term as in Eq.~(\ref{eq:ScalarEOM}).

 \subsection{Wigner space and gradient expansion}
 \label{subsec:Wigner}

% RELATE THIS TO NOTIONS OF MARKOVIAN/NONMARKOVIAN, DEFINE WHAT WE MEAN BY THIS
% STRESS IMPORTANCE OF  LINEAR APPROXIMATION WHERE IT APPLIES
% LITERATURE DISCUSSION

It is apparent that Eq.~\eqref{eq:ScalarEOM} describes the effects of other degrees of freedom on the evolution of the scalar field and that these are encoded in the self energy. However, the equation in this form is still a little formal in several aspects. Firstly, it is integro-differential which is impractical from the point of view of finding analytic solutions. Secondly, the self-energy is currently expressed in configuration space whilst we would intuitively expect information about the effects of a fermionic medium on the scalar to be expressible in terms of microphysical scattering processes, according to some particle or quasiparticle approximation. Indeed, we would hope to make quantitative predictions by calculating such processes and identifying them with $S$-matrix elements. Finally, this equation in its present form does not yet bear much resemblance to the usual form for a damped scalar field.  It turns out that all three of these issues can be solved by introducing the notion of a local (for each space-time point) momentum space, as captured by the \textit{Wigner transformation} defined by
% it appears to be advantageous to carry out the computations in Wigner space:\\
% To maintain the generality of the CTP approach, a representation in configuration space may be preferred.
 \begin{align}
 \label{Wigner:trafo}
     \Pi^R(q,x)=\int d^4 r\, e^{i q\cdot r}\, \Pi^R\left(x+\frac r2,x-\frac r2\right).
 \end{align}
%  where $A$ stands for a generic two-point function.
%  In particular, it may correspond to a correlation function or to a self energy.
% Note that in the limit where the background is translation invariant, 2-point functions depend only on the relative coordinate $\Pi^R(x_1,x_2) = \Pi^R(x_1-x_2)$ becomes exactly the Fourier transform. Thus we recognise the Wigner transform as a local momentum transformation with $x$ describing how the medium properties vary throughout spacetime.

 The approximations will rely on the mild dependence of $\Pi^R(q,x)$ on the average coordinate $x$. This is usually given by the macroscopic dependence on the background that is characterized by small derivatives $\partial_x^\mu$ with respect to $x$. On the other hand, the dependence on the relative coordinate corresponds to the quantum-physical correlations and is of order of the intrinsic energy-momentum scales given by $q^\mu$. Provided these generic assumptions hold, we can carry out a gradient expansion in terms of $q\cdot\partial_x$.
 
 Now, with this Wigner space description, we shall derive the damping term in the classical wave equation from collision integrals in a medium, as advertised in the Introduction. This connection appears as something that one would very much expect intuitively, and certainly it has been applied in one form or another in many instances. Nonetheless, we are not aware of a thorough derivation, which is what we shall present now.
 
 The Wigner transformation allows us to write
\begin{align}
& \int d^4 y \, \Pi^R(x,y)\phi(y)\notag\\
&=\int d^4 y \int\frac{d^4 q}{(2\pi)^4}\Pi^R\left(q,\frac{x+y}{2} \right)\phi(y) e^{-iq\cdot(x-y)}\notag. \\
 \end{align}
% \textcolor{purple}{put in configuration space expressions for correlaors?}
% We we used the fact that variation of $\Delta_2$ with respect to $\Delta$ generates the self energy
%     \begin{equation}
%         \partial^2 \phi(x) + \Pi^{\rm TP}(x) \phi(x) +
%     \int dy \Pi^R(x,y) \phi(y) = 0.
%     \end{equation}
%    Upon performing a Wigner transformation, defined by
%    \begin{equation}
%        \Pi(p,x) = \int dy \Pi \left(x + \frac{y}{2}, x- \frac{y}{2}\right) e^{ i y\cdot p}
%    \end{equation}
    % The main complication arises from the convolution term
    % \begin{align}
    %     &[\Pi^R\odot\phi](x)=\int d^4 y \Pi^R(x,y)\phi(y)
    %     \label{convolution:phi},
    % \end{align}
    % which implies that the wave equation~(\ref{eq:ScalarEOM}) is integro-differential.
    Our goal is to use the gradient expansion in order to reexpress this as a series of local terms so that we obtain an ordinary differential equation through its truncation.
    % Using the Wigner transform~(\ref{Wigner:trafo}), this can be written as
    % \begin{equation}
    %       [\Pi^R\odot\phi](x)=\int d^4 y \int\frac{d^4 p}{(2\pi)^4}\Pi^R\left(p,\frac{x+y}{2} \right)\phi(y) e^{-ip\cdot(x-y)}\notag\\
    % \end{equation}
    This is achieved through the Taylor expansion of $\Pi^R(q,(x+y)/2)$ about $q=0$. After repeated use of integration by parts (see Appendix \ref{App:Convolution}), one obtains the following expression for the convolution
    % Here, we have expressed $\phi(x-y)$ as a Taylor expansion about $\phi(x)$ in the third step, and in the fourth, we have integrated partially.
    % When the background can be approximated as homogeneous, the Wigner function $\Pi^R(p,x)\equiv\Pi^R(p)$ is independent of $x$ so that the convolution is
    \begin{align}
       &\int d^4 y \,\Pi^R(x,y)\phi(y) \nonumber \\
       & = \left[e^{i \partial_q \cdot \partial_y} \Pi^R\left(q,\frac{x+y}{2}\right) \phi(y)\right]_{q=0, y=x}\,.
        \label{convolution:homogeneous}
    \end{align}

    % The interpretation of this is simple: ...

    % However, in the present case, the background is not homogeneous because of the star being an object of finite extent. While we will use the homogeneous approximation for the self energy eventually, we further recast the full expression~(\ref{convolution:phi}) so that this procedure can be justified. For this purpose, we define the Fourier transform of the Wigner function with respect to the average coordinate
    % \begin{align}
    %     \widetilde\Pi^R(p,P)=\int d^4x e^{iP\cdot x}\Pi^R(p,x)\,.
    % \end{align}
    % Then, the convolution can be expressed as
    % \begin{align}
    %     [\Pi^R\odot\phi](x)=\int\frac{d^4 P}{(2\pi)^4}e^{-iP\cdot x}\widetilde\Pi^R(p,P)e^{i\overleftarrow{\partial_p}\cdot\overrightarrow{\partial_x}}\phi(x)\big|_{p=\frac{P}{2}}\,.
    % \end{align}
    % It can be seen immediately that in the homogeneous case, where $\Pi^R(p,P)=(2\pi)^4\delta^4(P)\Pi^R(p)$, Eq.~(\ref{convolution:homogeneous}) is recovered from this expression.

   This allows us to express Eq.~\eqref{eq:ScalarEOM} terms of the Wigner-transform of the self energy as
    \begin{equation}\label{eq:PiDiamond}
        \partial^2 \phi + \mu^2 \phi + 
    \left[e^{i \partial_q \cdot \partial_y} \Pi^R \left(q,\frac{x+y}{2}\right) \phi(y)\right]_{q = 0,x=y} \! \! \! = 0.
   \end{equation}
   Instead of an integro-differential equation, this is now in the form of an ordinary differential equation of infinite order.

 We note that for a real field $\phi$, ${\rm Re}[\Pi(q,x)]$ is an even function under point reflections in the four momentum $p$, while the spectral self energy
   \begin{align}\label{eq:PiA}
       \frac{i}{2}\left(\Pi^>(q,x)-\Pi^<(q,x)\right)=-{\rm Im}[\Pi^R(q,x)]
   \end{align}
   is odd. Therefore, Eq.~(\ref{eq:PiDiamond}) can be written as
   \begin{align}
       &\partial^2 \phi + \mu^2 \phi \nonumber \\
    +&\left[\cos(\partial_{q}\cdot\partial_{y}) {\rm Re}\left[\Pi^R \left(q,\frac{x+y}{2}\right)\right] \phi(y)\right]_{q = 0,x=y}\notag\\
    -&\left[\sin(\partial_{q}\cdot\partial_{y}) {\rm Im}\left[\Pi^R \left(q,\frac{x+y}{2}\right)\right] \phi(y)\right]_{q = 0,x=y} \nonumber \\
    =& 0.
    \label{eq:Pi:cos:sin}
   \end{align}
   In this form, it becomes explicit that the gradient expansion and Wigner space representation of the self energy respect that the field $\phi$ is real. Furthermore, the dispersive effects from ${\rm Re}[\Pi^R]$ and the absorptive ones from ${\rm Im}[\Pi^R]$ are thus separated.
   
   The wave equation for $\phi$  as in Eqs.~(\ref{eq:PiDiamond},\ref{eq:Pi:cos:sin}) is still non-Markovian because it is equivalent with Eq.~(\ref{eq:ScalarEOM}). In general, the non-Markovian property of an integro-differential equation implies a Wigner space description with an infinite series of derivative operators. In many different situations, it is however possible to truncate or to resum the derivatives so that one arrives at an effectively Markovian description. This is because the boundary data for a differential equation of finite order is fully specified by the field and an finite number of its derivatives at a given time. A typical approximation is to neglect the gradients $\partial_y\Pi^R$ when these correspond to macroscopic scales and are thus small compared to the scales that determine $\partial_p\Pi^R$, i.e. the density or temperature of the medium.
   
   \subsubsection{Damping in the long wavelength limit and superradiance}
   
   The gradients $\partial_y \phi$ may or may not be small. A generic situation is however $(\partial_q \Pi^R)\cdot(\partial_y \phi)\ll\Pi^R \phi$. This typically occurs when the wave number $|\mathbf k|$  and frequency $|k^0|$ (corresponding to $\partial_y$) are small compared to the density or temperature  $T$ of the medium (corresponding to $\partial_q$)---which is the case in the present application to stellar superradiance. Then, from Eq.~(\ref{eq:Pi:cos:sin}) it can be seen how damping enters the picture. When neglecting all derivatives in $\partial_q\cdot \partial_y$ but the leading one,
   %Now if the linear approximation ${\rm Im}\Pi^R(p,x)=p^\mu \left(\partial_{p^\mu}{\rm Im}[\Pi(p,x)]\right)_{p=0}$ holds and all higher derivatives with respect to $p$ can be neglected  acting on ${\rm Im}\Pi^R(p,x)$
   the contribution involving ${\rm Im}[\Pi^R]$ in Eq.~(\ref{eq:Pi:cos:sin}) takes the form of a classical friction term in covariant form:
\begin{align}\label{eq:partialphi}
     \partial^\mu_q {\rm Im}\left[\Pi^R(q,x)] \right]_{q=0}\, \partial_\mu \phi.
   \end{align}
   We typically envisage a situation in which the damping arises as the result of the absorption of a soft $\phi$ mode onto the external leg of an existing in-medium scattering process---see Fig.~\ref{fig:MatrixElts}. In this case, the leading four-momentum transfer into the medium comes from $q^0$ component whilst three-momentum shift $\textbf{q}$ is sub-leading, so that the axion absorption is elastic with respect to the three-momenta of the medium particles. This allows us to neglect spatial momenta in $\Pi^R$ to leading order.  This can be seen if one takes a diagram in Fig.~\ref{fig:MatrixElts}, where the axion momentum enters via the nucleon propagator $1/[(p+q)^2 - m^2] = 1/[2p^0 q^0 - 2\textbf{p}\cdot \textbf{q} + q^2]$, since $p^2 = m^2$ where $m$ and $p$ is the mass of an in-medium particle. Clearly for non-relativistic particles, $p^0 \simeq m \gg |\textbf{p}|$ so that the process is dominated by $q^0$ with the $\textbf{q}$ dependence being suppressed by the heavy mass of the medium particles. As a result we can approximate
   \begin{equation}
       \Pi^R(q^0,\textbf{q},x ) \simeq \Pi^R(q^0,0,x).
   \end{equation}
   This means that the spatial derivative term in  Eq.~\eqref{eq:partialphi} is effectively suppressed.
Then, since ${\rm Im}[\Pi^R]$ is odd in $q^0$ it follows that we can derive an effective damping rate
   \begin{align}
       \Gamma_\phi &=  - \left. \partial_{q_0}\text{Im} \Pi^R(q^0,0,x)\right|_{q^0\rightarrow 0} \nonumber \\
       &= - \lim_{q^0\rightarrow 0}\frac{\text{Im} \, \Pi^R(q^0,0, x)}{q^0}.
   \end{align}
   leading to an equation
%   We can then define a microphysical damping rate $\Gamma_\phi(p)$ defined by
%   \begin{equation}
%       \Gamma_\phi(p) = \frac{\text{Im} \, \Pi^R(p, x)}{p^0}.
%   \end{equation}
%   Note therefore that $p^0$ breaks the symmetry between $p^0$ and $\textbf{p}$ so that spatial derivatives in \eqref{eq:partialphi} are higher order in $p^\mu$, so that to leading order one has only a temporal derivative, leading to
   \begin{equation}\label{eq:DampedScalar}
       \partial^2 \phi  + \mu^2 \phi + \Gamma_\phi \dot{\phi} = 0.
   \end{equation}

   \subsubsection{Damping in the short wavelength limit (WKB)}

   For completeness of this discussion, we also consider the case when the gradients $\partial_y \phi$ are not small, i.e. when the wave number $|\mathbf k|$ or frequency $|k^0|$ of the field is of the same order or large compared to the temperature $T$ or density. Another setup where these gradients cannot be neglected is when the wave propagates in vacuum and the invariant mass of $\phi$ is such that the decay into two or more particles is possible. The corresponding momentum thresholds in ${\rm Im}[\Pi^R]$ can then invalidate the truncation of the gradient expansion. For comparison, note that for the convolution of two-point functions in Wigner space, large derivatives because of large wave numbers or frequencies do not occur because these are accounted for by the momentum variables in Wigner space. However, as we shall see now, for approximate plane waves, the derivatives can be summed to all orders so that once again an effectively Markovian description results.
   
   In order to see that we recover in such situations from Eq.~(\ref{eq:Pi:cos:sin}) the standard corrections to the dispersion relations and the damping effects, assume for now that the background is spatially homogeneous and time-independent, i.e. $\Pi^R(q,x)\equiv \Pi^R(q)$. A relaxation of the  homogeneity conditions may be dealt with using the Wentzel--Kramers--Brillouin approximation (the applicability of which will depend on the specific problem).
   
   Then, the spatial part of the solutions can be expanded in a basis of momentum eigenstates $\phi(x)\sim\exp(i\mathbf k\cdot\mathbf x)$. When summing the spatial derivatives to all orders, we recognize a Taylor expansion of the self energy (note that this relies on the exponential form of the solution) so that Eq.~(\ref{eq:PiDiamond}) becomes
   \begin{align}
       (\partial_t^2+\omega^2)\phi(t,\mathbf x)+\left[e^{i\partial_{q^0}\partial_y^0}\Pi^R(q^0,\mathbf k)\phi(y^0)\right]_{q^0=0,y^0=t}=0,
   \end{align}
   where $\omega=(\mathbf k^2+m^2)^{1/2}$.
   
Neglecting a possible dependence of $\Pi^R$ on $\phi$, this equation is linear so that the solution is of the form $\phi(x)=\exp(-i\omega^{(1)}t+i\mathbf k\cdot\mathbf x)$, where the notation $\omega^{(1)}$ indicates that i corresponds to a correction to $\omega$ due to $\Pi^R$. Then, also the temporal gradients can be summed, what leads to
\begin{align}
    (\partial_t^2+\omega^2)e^{-i\omega^{(1)} t}+\Pi^R(\omega^{( 1)},\mathbf k)e^{-i\omega^{(1)}} t=0
\end{align}
   and therefore
   \begin{align}
      {\omega^{(1)}}^2=\omega^2 + \Pi^R(\omega,\mathbf k).
   \end{align}
  In the argument of $\Pi^R$, we have replaced $\omega^{(1)}$ with $\omega$. This leads to a difference at  second order in the self energy, while we are aiming here to consistently include the first order effects.
 
%   We thus see that....

% For long-wavelengths, we truncate the gradient expansion, retaining only the local piece of the self-energy
%  \begin{equation}
%         \partial^2 \phi + \Pi^{\rm TP} \phi +
%     \Pi^R(\omega,0, x) \phi(x) = 0
%     \end{equation}
% We should re-check if the dispersive part of $\Pi^R$ can be neglected relative to the absorptive part in light of the fact we know the spurious $(1+f_B(T))$ enhancement is gone. We can put the mass correction in an appendix.

% We can put some generic self energy diagrams with different topologies in and show their cuts?

%  \begin{equation}\label{eq:DampedScalar}
%         \partial^2 \phi + \Pi^{\rm TP} \phi +
%     \text{Im} \Pi^R(0, x) \phi(x) = 0
%     \end{equation}
%     This is equivalent to neglecting dispersive effects in the response of the medium. We shall see through explicit computation of $\text{Im} \Pi^R$ for axions, that such an approximation is justified in practice. The spectral self energy is related to the Wightman self-energies via
% \begin{equation}
%     \Pi^> = \Pi^{-+}, \qquad   \Pi^< = \Pi^{+-}
% \end{equation}
% in the form
% \begin{align}\label{eq:PiA}
%     \Pi^{\cal A}(\omega ) & =
%     \frac{i}{2}\left( \Pi^>(\omega) -\Pi^<(\omega)  \right) \nonumber  \\
%     & =\frac{i}{2} \Pi^>(\omega) \left( 1 - e^{- \omega / T }\right)  
% \end{align}
% where we used the KMS relation $\Pi^> = e^{\omega/T}\Pi^<$.

\section{Superradiance with worldline effective field theory}\label{sec:EFT}

% \textit{Contributor: Fran Chadha-Day}

To describe stellar superradiance, we are interested in the interaction of our field with a {\it rotating} star. One strategy would be to attempt to include rotation in the formalism described in Section \ref{sec:NeqQFT} by trying to introduce a bulk azimuthal fluid velocity into the distribution functions. It remains to be seen how this would be realised in detail. Instead, in this work, we will make use of the worldline effective field theory  (EFT) approach developed in \cite{Endlich:2016jgc}. This will simplify the calculations and also gives insight into the physics of superradiance. This provides a relation between the absorption of modes into a static non-rotating star and the amplification of modes into a rotating star, with the transformation being inferred from an appropriate transformation between a rotating and non-rotating frame.

As with all EFTs we must exploit a hierarchy of scales. In the present case, the high energy scale is set by the radius of the star or black hole $\sim R$, and the low energy scale corresponds to wavelengths $\sim 1/k$ of the superradiant field which are large  compared to this scale. The EFT is then an effective expansion in the parameter $R/k^{-1} \ll 1$.

% with the star being effectively ``integrated out" so that the infinite tower of operators describing contact interactions between the field and the star.  

% A worldline EFT describes the physics of an extended object whose size is much smaller than other relevant length scales (in this case the Compton wavelength of the field).

In this picture, the extended object is essentially integrated out, leaving a power series of contact interactions between an effective point-like object and the field around it. In this way, finite size effects are then described by an infinite series of effective operators that act in the worldline of the object. The symmetries of these operators are determined by the symmetries of the object. As described in \cite{Delacretaz:2014oxa}, a spherically symmetric object is described by operators transforming under $SO(3)$.

For example, for an extended object interacting with a scalar field $\phi$, the interaction Hamiltonian in the worldline EFT is

\begin{align}\label{eq:Hamiltonian}
    H_{\rm int}(t,{\bf x}) = & \partial^{I} \phi(x) \mathcal{O}^{(1)}_{I}(x) \delta^{(3)}({\bf x} - {\bf y}(t)) + \\ & \partial^{I} \partial^{J} \phi(x) \mathcal{O}^{(2)}_{I J}(x) \delta^{(3)}({\bf x} - {\bf y}(t)) + \dots, \nonumber
\end{align}
where ${\bf y}(t)$ is the worldline of the extended object and $\mathcal{O}^{(n)}$ are effective operators describing the interaction between the object and the scalar field. $I, J$ are {\it spatial} indices, so that the operators $\mathcal{O}^{(n)}$ are not Lorentz invariant. This arises because a macroscopic object breaks Lorentz symmetry through its rest frame, location and orientation in space.

The worldline EFT has many applications, including in black hole superradiance \cite{Endlich:2016jgc}. One powerful aspect is the ease with which rotation of the extended object can be added. Firstly, we can integrate the Hamiltonian over space to project the Hamiltonian onto the worldline of the star:
\begin{align}
    H_{\rm int}(t) & = \int d^3 {\bf x} H_{\rm int}(t,{\bf x}) \nonumber \\ & =  \partial^{I} \phi(t)
    \mathcal{O}^{(1)}_{I}(t) +  \partial^{I} \partial^{J} \phi(t)  \mathcal{O}^{(2)}_{IJ}(t) + ... ,
\end{align}
where the fields are now understood to be evaluated at the spatial position of the object which we will take to be ${\bf x} = 0$. The EFT associated to a rotating object can be inferred through the action of a time-dependent rotation matrix on the operators $\mathcal{O}^n$
\begin{equation}
\mathcal{O}^{(n)}_{I_1 \cdots I_n}(t) \rightarrow  \tensor{R}{^{J_1} _{I_1}}(t)\cdots \tensor{R}{^{J_\ell} _{I_n}}(t) \mathcal{O}^{(n)}_{J_1 \cdots J_n}(t),
\end{equation}
where, without loss of generality we have chosen coordinates in which the object spins about the $z$-axis so that
\begin{equation}
R(t)=\left(\begin{array}{ccc}
\cos (\Omega t) & -\sin (\Omega t) & 0 \\
\sin (\Omega t) & \cos (\Omega t) & 0 \\
0 & 0 & 1
\end{array}\right).
\end{equation}
The EFT for the field interacting with a rotating object, as as seen by an inertial (non-rotating), observer then reads
    \begin{align}\label{eq:rotatedH}
    H_{\rm int}(t) &  =  \partial^{I} \phi(t) \tensor{R}{_I ^J}(t) \mathcal{O}^{(1)}_{J}(t) \nonumber \\
    &+  \partial^{I} \partial^{J} \phi(t) \tensor{R}{_I^K}(t) \tensor{R}{_J ^L}(t) \mathcal{O}^{(2)}_{K L}(t) + ... ,
\end{align}
% As shown in \cite{Delacretaz:2014oxa}, the field should be transformed to the instantaneous rest frame of the object.

% Assuming the boost to the rest frame has already been taken care of, the transformation is given simply by applying a time-dependent rotation $R_a^b(t)$ about the spin-axis.
% \begin{align}\label{eq:rotatedH}
%     H_{\rm int}(t) & = \int d^3 {\bf x} H_{\rm int}(t,{\bf x}) \nonumber \\ & =  \partial^{I} \phi(t) \tensor{R}{_I ^J}(t) \mathcal{O}^{(1)}_{J}(t) \nonumber \\
%     &+  \partial^{I} \partial^{J} \phi(t) \tensor{R}{_I^K}(t) \tensor{R}{_J ^L}(t) \mathcal{O}^{(2)}_{K L}(t) + ... ,
% \end{align}
% We choose coordinates in which the object rotates about the $z$ axis with a constant angular momentum $\Omega$, such that

% MAYBE PUT A NOTE EXPLAINING CORRELATORS ARE KEY! AND EXPLAIN THE $\rho(\omega < m_a)$ and $\rho(\omega > m_a)$ arguments and how massless scattering contains enough information to extract the sub-mass spectral information. You continue the spectral function to values of $m_a \rightarow 0$ and that allows you to extract enough information to get bound states.

In the following sections we will show how this EFT can be used to describe superradiant scattering and superradiant instabilities of bound modes around a rotating object.  

\subsection{Superradiant scattering}\label{sec:SuperradiantScattering}

First, let us imagine the interaction Hamiltonian is switched on over some finite time $\mathcal{T}$. The evolution operator is then given by

\begin{align}\label{eq:EvolutionOp}
S & = T \exp \left\{-i \int_0^\mathcal{T} dt \int d^3{\bf x}  H_{\mathrm{int}}(t, {\bf x})\right\} \\& = T \exp \left\{-i \int_0^\mathcal{T} d t H_{\mathrm{int}}(t)\right\}.
\end{align}

Let us now consider the effect of this interaction on the absorption of the field $\phi$ by the rotating object, following the method introduced in \cite{Endlich:2016jgc}.  We compute the absorption of a spherical mode of the field $\phi$ with frequency $\omega$ and angular momentum labels $\ell,m$. The probability for this process is:
\begin{equation}\label{eq:probability}
    P_{\mathrm{abs}}=\sum_{X_{f}} \frac{\left|\left\langle X_{f} ; 0|S| X_{i} ; \omega, \ell, m\right\rangle\right|^{2}}{\langle\omega, \ell, m \mid \omega, \ell, m\rangle},
\end{equation}
where $X_i$ and $X_f$ are the initial and final states of the star.  Note that the leading order contribution to scattering of a mode with angular momentum $\ell$ comes from the operator
\begin{equation}\label{eq:Hell}
  H_{\ell}(t) =  \partial^{I_1}\cdots \partial^{I_\ell} \phi \tensor{R}{_{I_1} ^{J_1}}\cdots \tensor{R}{_{I_\ell}^{J_{\ell}}}  O_{J_1...J_\ell},
\end{equation}
in Eq.~\eqref{eq:Hamiltonian}.
% with $\ell$ derivatives contributes to scattering of radiation with angular momentum quantum number $\ell$.
% We emphasise that in spite of the suggestive notation, eqs.~\eqref{eq:Hamiltonian} and \eqref{eq:probability} are entirely classical. The above formalism describes an effective Hilbert space of classical modes which interact with the star. The notation in Eq.~\eqref{eq:probability} is just a short-hand to describe a complete basis of classical modes, on which one can define some inner product. These are \textit{not} quantum mechanical eigenstates in any sense. \textcolor{red}{Perhaps we should explain this further or take it out?}
Upon inserting \eqref{eq:Hell} into \eqref{eq:EvolutionOp} and expanding to leading order in interactions, one obtains
\begin{align}
    & \sum_{X_{f}} \left|\left\langle X_{f} ; 0|S| X_{i} ; \omega, \ell, m\right\rangle\right|^{2}
    =  \nonumber \\
    & \int_0^\mathcal{T} dt \int_0^\mathcal{T} dt' \nonumber \\
    & \bra{0}\partial^{I_1}\cdots \partial^{I_\ell} \phi(t') \tensor{R}{_{I_1} ^{J_1}}(t')\cdots \tensor{R}{_{I_\ell}^{J_{\ell}}}(t')\ket{\omega,\ell,m} \nonumber \\
    & \bra{\omega,\ell,m} \partial^{K_1}\cdots \partial^{K_\ell} \phi(t) \tensor{R}{_{K_1} ^{L_1}}(t)\cdots \tensor{R}{_{K_\ell}^{L_{\ell}}}(t) \ket{0}
    \nonumber \\
    &\braket{O_{J_1...J_\ell}(t^{\prime}) O_{L_1...L_\ell}(t)},
\end{align}
where  $\braket{O_{J_1...J_\ell}(t^{\prime}) O_{L_1...L_\ell}(t)}$ denotes the expectation value for the initial state $|X_{i}\rangle$. All that remains is to manipulate the matrix elements associated to the spherical modes of $\phi$. This is a lengthy but straightforward computation, the details of which can be found in \cite{Endlich:2016jgc}. The result is
\begin{align}\label{eq:MatrixAbsElt}
& \sum_{X_{f}} \left|\left\langle X_{f} ; 0|S| X_{i} ; \omega, \ell, m\right\rangle\right|^{2} \\
&=\frac{\ell ! q^{2 \ell + 2}} {2 \pi v \omega (2 \ell +1 ) !!} \int \frac{d \omega^{\prime}}{2 \pi}  \Delta_{\ell}(\omega')\left| \int_0^\mathcal{T} dt {\rm e}^{i \omega^{\prime} t} {\rm e}^{-i (\omega - m \Omega) t} \right|^2,
\end{align}
where $q = |\textbf{q}|$ and $v = \sqrt{1 - \mu^2/\omega^2}$ is the group velocity of a field mass $\mu$ asymptotically far from the star and $\Delta_{\ell}(\omega)$ is the Fourier transform of the Wightman correlation function defined by
% $\langle \mathcal{O}_a(t) \mathcal{O}_b(t^{\prime}) \rangle$:
% \begin{equation}
%     \left\langle\mathcal{O}_{J}\left(t^{\prime}\right) \mathcal{O}_{L}(t)\right\rangle=\delta_{J L} \int \frac{d \omega^{\prime}}{2 \pi} \Delta\left(\omega^{\prime}\right) e^{i \omega^{\prime}\left(t-t^{\prime}\right)}.
% \end{equation}
\begin{equation}
    \braket{O_{J_1...J_\ell}(t^{\prime}) O_{L_1...L_\ell}(t)} = \delta_{J_1...J_\ell}^{L_1...L_\ell} \int \frac{d \omega}{2 \pi} \Delta_\ell(\omega) {\rm e}^{i \omega (t - t^{\prime})}.
\end{equation}
The above form is permitted by isotropy (spherical symmetry of the star), leading to delta function prefactors, and the fact the background is stationary so that time-translation symmetry means the correlators can only dependent on the relative coordinate $t - t'$. For sufficiently long times the $t$ integral can be approximated by a delta function, multiplied by $\mathcal{T}$ leading to
\begin{align}
&\sum_{X_{f}} \left|\left\langle X_{f} ; 0|S| X_{i} ; \omega, \ell, m\right\rangle\right|^{2} \nonumber \\
& =\frac{\ell ! q^{2 \ell + 2}} {2 \pi v \omega (2 \ell +1 ) !!} \int \frac{d \omega^{\prime}}{2 \pi}  2 \pi \mathcal{T} \delta_{\ell}(\omega^{\prime} - (\omega - m \Omega)) \Delta_\ell(\omega') \nonumber \\& = \frac{\ell ! q^{2 \ell + 2}} {2 \pi v \omega (2 \ell +1 ) !!} \Delta_{\ell}(\omega-m \Omega) \mathcal{T}. 
\end{align}
%  We will follow the usual procedure of dividing by $\mathcal{T}$ to obtain a rate.
 We now turn to the normalisation of states in the denominator, which, as we show in appendix \ref{App:Scattering} is normalised to \footnote{Note that our treatment of limits in deriving the absorption and emission rates differs slightly from that in \cite{Endlich:2016jgc}. In \cite{Endlich:2016jgc}, the $\delta(0)$ terms corresponding to the infinite time and infinite radius limits are incorrectly canceled. This error does not affect their conclusions.}
 \begin{equation} \label{eq:denominator}
     \braket{\omega \ell m \left| \omega \ell m \right. } = \frac{2 \mathcal{R}}{v},
 \end{equation}
 where $\mathcal{R}$ is a finite spherical radius used to regulate the inner product of non-normalised states. This is the spherical equivalent of the usual replacement   $(2 \pi)^3 \delta^{(3)}({\bf k}=0) \rightarrow V$.
Hence we can write
\begin{equation}\label{eq:Pabs}
    P_{\rm abs} = \frac{\ell ! q^{2 \ell + 2}} {4 \pi  \omega (2 \ell +1 ) !!} \Delta_{\ell}(\omega-m \Omega) \frac{\mathcal{T}}{ \mathcal{ R}}.
\end{equation}
% (----------)\\
% % The Wightman correlation function is given by $\Delta_{\ell}(\omega) = \theta(\omega)\rho_{\ell}(\omega)$, where $\rho_{\ell}(\omega)$ is the spectral density.
% We can expand $\rho_{\ell}(\omega)$ as $\rho_{\ell}(\omega) = \gamma_{\ell} \omega+ \mathcal{O}(\omega^3)$, where $\gamma_{\ell}$ is a constant.
Dividing by $\mathcal{T}$, we then obtain an absorption \textit{rate}
% \begin{equation}
% \label{eq:absorptionRate}
%     R_{\rm abs} = \theta(\omega - m \Omega)\frac{k^{4} \delta_{1}^{l}}{6 v} \gamma_1 (\omega-m \Omega) n_r,
% \end{equation}
\begin{equation}
\label{eq:absorptionRate}
    R_{\rm abs} = \frac{\ell ! q^{2 \ell + 2}} {4 \pi \omega (2 \ell +1 ) !!} \Delta_{\ell}(\omega- m \Omega) n_r,
\end{equation}
where $n_r= 1/\mathcal{R}$ is the particle number per spherical shell of unit radius. We can similarly calculate an emission probability
\begin{equation}
    R_{\rm em} = \frac{\ell ! q^{2 \ell + 2}} {4 \pi \omega (2 \ell +1 ) !!} \Delta_{\ell} (m \Omega - \omega) n_r.
\end{equation}
% \begin{equation}
%     P_{\rm em} = \frac{\ell ! k^{2 \ell + 2}} {4 \pi  \omega (2 \ell +1 ) !!} \Delta_{\ell} (m \Omega -\omega) \frac{\mathcal{T}}{ \mathcal{ R}},
% \end{equation}
noting that the argument of $\Delta_\ell$ appears now with the opposite sign to Eq.~\eqref{eq:Pabs}.
% We see that absorption is non-zero only for $\omega > m \Omega$ while for $\omega < m \Omega$ the rotating body emits radiation i.e. we have superradiance scattering.
We can then define an amplification factor for spherical waves as
\begin{align} \label{eq:Zlm}
Z_{\ell m} &=
% \lim_{\mathcal{T}/\mathcal{R} \rightarrow 1} \left[ P_{\rm em} - P_{\rm abs}\right] \nonumber \\
\frac{\Phi_{\rm out}- \Phi_{\rm in}}{\Phi_{\rm in}} \nonumber \\
&= \frac{R_{\rm em} - R_{\rm abs}}{n_r v}\nonumber \\
&=  \frac{\ell! \, q^{2 \ell + 2}}{4 \pi (2 \ell+1)!! v \omega} \, \rho_\ell (m \Omega - \omega ),
\,
% \\ & \simeq -\frac{\ell! \, k^{2 \ell + 2}}{2 \pi (2 \ell+1)!! v \omega} \, \gamma_\ell (\omega - m \Omega). \nonumber
 \end{align}
where $\Phi_{\rm in} = n_r v$ is the ingoing flux, which is related to the outgoing flux by $\Phi_{\rm out} = \Phi_{\rm  in}[1 + P_{\rm em} - P_{\rm ab}]$ and where
\begin{equation}
    \rho_\ell(\omega) = \Delta_\ell(\omega) - \Delta_\ell(-\omega),
\end{equation}
is the spectral response function for the star. The key point is that $\rho_\ell(\omega)$ is an odd function of $\omega$, hence for frequencies $\omega > m \Omega$ energy is absorbed ($Z_{\ell m}<0$) whilst for $\omega > m \Omega$  energy is extracted from the star ($Z_{\ell m}>0$). This is superradiant scattering.

\subsection{Superradiant instabilities} \label{sec:BoundStates}

We now turn to a similar question of bound states, extending the method of \cite{Endlich:2016jgc}. It is precisely these bound states, which, due to their gravitational confinement, exhibit instabilities through the superradiant extraction of rotational energy. Consider bound states of a spin-0 field with mass $\mu$. The scalar field operator can be rewritten in terms of bound state creation and annihilation operators:

\begin{align}\label{eq:ModeExpansion}
    \hat{\phi} = \sum_{n \ell m}  \frac{1}{ \sqrt{2 \omega_{n\ell}} }
    \Big( &\hat{a}_{n \ell m} f_{n\ell m} (r, \theta, \phi) {\rm e}^{-i \omega_{n \ell} t} \nonumber \\  &+\hat{a}^{\dagger}_{n\ell m} f^{\star}_{n\ell m} (r, \theta, \phi) {\rm e}^{i \omega_{n \ell} t} \Big),
\end{align}
where $a_{n \ell m}$ are creation operators for a mode $(n , \ell,  m)$. The mode functions
\begin{equation}
f_{n \ell m}(r,\theta, \phi) = R_{n \ell}(r)Y_{\ell m}(\theta, \phi),
\end{equation}
are the bound state solutions on a Schwarzschild black-hole background. Expressions for $R_{n \ell}$ are given in appendix \ref{App:BoundStates}.
% solves the Hydrogen-like Schrodinger equation for the gravitational attraction between the particle and the star,
They are normalised as:
\begin{equation}
    \int d \Omega dr r^2 f_{n \ell m}(r,\theta,\phi) f^{\star}_{n^{\prime}\ell^{\prime}m^{\prime}}(r, \theta, \phi) = \delta_{n n^{\prime}} \delta_{\ell \ell^{\prime}} \delta_{m m^{\prime}}.
\end{equation}
$n$ and $l$ take non-negative integer values and $m$ takes integer values from $-l$ to $l$.  
We wish to calculate the absorption and emission probability of a bound state $\ket{n,\ell,m} = \hat{a}^{\dagger}_{n\ell m} \ket{0}$ by the star. We have:
\begin{equation}
P_{\rm abs} = \sum_{X_f} | \bra{X_f; 0} S \ket{X_i; n \ell m}|^2,
\end{equation}
where $X_i$ and $X_f$ are the initial and final states of the star. Note that the bound states are normalised to 1. As before, the dominant contribution to this process comes from the interaction of Eq.~\eqref{eq:Hell}, leading to
% $S$ is derived from the interaction Hamiltonian for a given multipole $l$:
% \begin{equation}
%     H_{int} = \partial^{I_1}...\partial^{I_\ell} \phi R_{I_1}^{J_1}...R_{I_\ell}^{J_\ell} O_{J_1....J_\ell},
% \end{equation}
% where capital letters represent spatial indices, $R$ are rotation matrices and $O$ is the composite operator parameterising finite size corrections along the star's worldline. We then have:
% \begin{equation}
%     S = T {\rm exp} \left( - i \int dt H_{int}(t) \right).
% \end{equation}

% We are working in the rest frame of the star, and hence the star's worldline corresponds to an integral over $t$ only. Expanding $S$ to first order in the interaction Hamiltonian, we have:
% As before the dominant contribution to this process comes from the operator \eqref{eq:Hell} which gives
\begin{align}
    P_{\rm abs} \simeq \sum_{X_f} \int^{\mathcal{T}}_0 & dt dt^{\prime} \bra{X_i; n,\ell,m} H_{\ell}(t^{\prime}) \ket{X_f;0} \\& \bra{X_f; 0} H_{\ell}(t) \ket{X_i;n,\ell,m}. \nonumber
    \label{eq:PabsBoundState}
\end{align}
Whilst the general $\ell$ case for superradiant scattering has been covered in \cite{Endlich:2016jgc}, to which we referred the reader earlier in subsection \ref{sec:SuperradiantScattering}, the general $\ell$ bound state case was not presented there. We therefore collect the relevant details in Appendix \ref{App:BoundStates}. There, we show that for large $\mathcal{T}$, one finds
%\begin{align}
 %   &P_{\rm abs} = A_{n \ell m} \left( \frac{2}{a n} \right)^{2\ell +3} \int \frac{d \omega}{2 \pi} \frac{\Delta_\ell(\omega)}{2 \mu} [(2 \pi)^2 \delta(m \Omega + \omega - \mu)]^2,
% &   {\normaltext \text{\quad \textcolor{purple}{Fran Fix}} }
%\end{align}
\begin{align}
    &P_{\rm abs} = \nonumber \\
    &A_{n \ell m} \left( \frac{1}{r_{n \ell}} \right)^{2\ell +3} \int \frac{d \omega}{2 \pi} \frac{\Delta_\ell(\omega)}{2 \omega_{\ell n}} \mathcal{T} (2 \pi) \delta(m \Omega + \omega - \omega_{n\ell}),
% &   {\normaltext \text{\quad \textcolor{purple}{Fran Fix}} }
\end{align}
% \textcolor{purple}{why here do you just write something with the delta function rather than a normalised object?}
where $r_{n\ell} = (n+\ell+1)/(2GM \mu^2)$ and $\omega_{\ell n} $ are the discrete bound state frequencies defined in Eq.~\eqref{eq:BSFreqs} and
%\begin{align}
 %   A_{n \ell m} & = \frac{l!(2\ell +1)!!}{8 \pi n (n-l-1)! (n+l)! (2\ell +1)!},
%\end{align}

% \begin{align}
%     A_{n \ell m} & = \frac{l!(2\ell +1)!!(n+l)!}{8 \pi n (n-l-1)! (2\ell +1)!^2}, (FRAN)
% \end{align}
\begin{equation}
    A_{n \ell m} = \frac{\ell! (2 \ell+1)\text{!!} (2 \ell+n+1)!}{8 \pi  (\ell+n+1) n! (2 \ell+1)!^2 }. 
\end{equation}
%The absorption rate is the absorption probability per unit time which we can write as $\Gamma_{\rm abs} 2 \pi \delta(0) = P_{\rm abs}$, following the same procedure as in the standard derivation of scattering cross sections.
We then obtain the absorption rate
\begin{equation}
    \Gamma_{\rm abs} = A_{n \ell m} \left( \frac{1}{r_{n \ell}} \right)^{2\ell +3} \frac{1}{2 \omega_{\ell n} } \Delta_\ell(\omega_{\ell n} - m \Omega).
\end{equation}
Similarly, the emission rate is
\begin{equation}
    \Gamma_{\rm em} = A_{n \ell m} \left( \frac{1}{r_{n \ell}} \right)^{2\ell +3} \frac{1}{2 \omega_{\ell n} } \Delta_\ell(m \Omega - \omega_{\ell n}),
\end{equation}
and so the superradiance rate is simply the difference of these two quantities
% \begin{equation}
%     \Gamma_{n \ell m} = A_{n \ell m} \frac{1}{2 \mu} \gamma_l (m \Omega - \mu),
%     \label{SRrate}
% \end{equation}
\begin{align}\label{eq:SRrate}
    \Gamma_{n \ell m} &=  \frac{A_{n \ell m}}{2 \omega_{\ell n} } \left( \frac{1}{r_{n \ell}} \right)^{2\ell +3} \rho_\ell(m \Omega - \omega_{\ell n} ).
    % \nonumber \\
    % &\simeq  \frac{A_{n \ell m}}{2 \mu} \left( \frac{1}{r_{n \ell}} \right)^{2\ell +3} \gamma_\ell(m \Omega - \mu ).
\end{align}
%This can then be related to the superradiant scattering rate given in equation \eqref{amplification} through
%\begin{equation}
 %   \Gamma_{n \ell m} =  \frac{A_{n \ell m}}{2 \mu} \frac{4 \pi (2 \ell+1)!! v \omega}{\ell! \, k^{2 \ell + 2}} Z_{\ell m}.
%\end{equation}

%where $a$ is the gravitational Bohr radius.

% All that remains now is to obtain the value of $\gamma_{\ell}$ for the star or black hole under consideration.

As we have seen in the previous section, and as described in \cite{Endlich:2016jgc}, we can obtain $\rho_\ell$ by matching to the superradiant scattering amplitude \eqref{eq:Zlm}, so that scattering off a static star leads immediately to the superradiant growth rate of bound states. By definition, gravitational bound states have frequencies $\omega <\mu$ below the mass of the field. Hence superradiance requires knowledge of $\rho_\ell(\omega < \mu )$ and since a massive propagating wave has $\omega >\mu$, it follows that carrying out scattering of massless modes with $\omega < \mu$ is sufficient to extract the spectral response at arbitrarily small arguments. This can then be used to infer the superradiance rate for massive fields which requires $\rho(\omega)$ in the region $\omega < \mu$. We illustrate this in the next subsection for a uniform star model.

The crucial point of this section is that correlators $\braket{O_{J_1...J_\ell}(t^{\prime}) O_{L_1...L_\ell}(t)}$ encode all the information about the response of a point-like star to a harmonic stimulus $\sim e^{- i \omega t}$ from the external field which is essentially applied homogeneously across the star in the long wavelength limit. Hence knowledge of these correlators, as embodied in $\rho$ is sufficient to calculate superradiance rates and this can be extracted from any physical process to which $\rho$ contributes.

%where we have used the results of section \ref{sec:NeqQFT} to equate their damping coefficient $\alpha$ with $\frac{{\rm Im}(\Pi^{\mathcal{A}}) }{\omega}$. \textcolor{red}{Note - it looks like there is a type in the original paper as $\Omega$ should be $m \Omega$. Since we are allowed to do the matching for a non-rotating star, this doesn't actually affect the result anyway.} By matching this with equation \eqref{amplification}, we can obtain the value of $\gamma_l$ and hence the superradiance rate for bound states from equation \eqref{SRrate}:

% \textcolor{red}{Change $a$ to standard notation !}
% \textit{Here fran can argue that it would have been messy to put curvilinear/azimuthal currents into the distribution functions for fermions above. Instead we can compute the superradiance rate for a static (non-rotating star) and then using the SO(3) rotating argument}

\subsection{Superradiance rate from microphysics}

% We must now obtain an expression for the worldline EFT coefficients $\gamma_l$ by a matching calculation. For example, the amplification of an incoming spherical wave, in the Newtonian and non-relativistic limit, found in \cite{1505.05509} is:  
As an example, let us consider superradiance for a scalar field around a uniform star. To obtain the spectral function $\rho$, we will calculate the scattering of a massless scalar field $\phi$ from the star when it is not rotating. For a uniform star the self-energy $\Pi^R$ is constant inside $r\leq R$ and zero elsewhere. The equation of motion for $\phi$ is given by equation \eqref{eq:DampedScalar}. Under these approximations, we can compute the absorption of a mode $(\ell,m)$ by considering solutions of the form
\begin{equation}
    \phi = \psi(r)  Y_{\ell m}(\theta,\varphi) e^{- i \omega t},
\end{equation}
which gives a radial equation
\begin{equation}
 r^2 \psi'' + 2 r \psi'   + \left[ r^2 ( \omega^2 + i\omega \Gamma_\phi  ) - \ell(\ell  + 1)\right]\psi =0.
\end{equation}
We recognise this immediately as the equation for spherical Bessels functions which admits two linearly-independent solutions. Inside the star, by regularity at the origin, the solution must take the form $j_\ell\left(r \sqrt{\omega^2 + i \omega \Gamma_\phi }\right)$ whilst outside it can be most conveniently parameterised in terms of spherical hankel functions $h_{\ell}^{(1,2)}(\omega r)$. Hence at infinity, the solution takes the form
\begin{equation}
    \psi(r\rightarrow \infty) = \phi_{\rm out} \frac{e^{i \omega r}}{r} + \phi_{\rm in} \frac{e^{-i \omega r}}{r}.
\end{equation}
Imposing continuity of the solution and its first derivative at $r=R$ gives expressions for $\phi_{\rm in}$ and $\rm \phi_{\rm out}$. One can then define an amplification factor for a non-rotating star in flat space:
\begin{align}
&Z_{\ell m }= \frac{|\phi_{\rm out}|^2}{|\phi_{\rm in}|^2} -1. 
\end{align}
Using the above continuity conditions, expanding in the limit $\omega R \ll 1$ and to linear order in $\Gamma_\phi $, we find (recalling we deal first with a non-rotating star)
\begin{equation}\label{eq:ZlmStar}
Z_{\ell m } \simeq -\frac{4 R^2   \omega (R \omega)^{2\ell + 1}}{(1+ 2\ell )!! (2\ell + 3 )!!}\cdot\Gamma_\phi
\end{equation}
repeating the expression found in \cite{Cardoso:2015zqa}.
% Computing these matrix elements involves similar steps to Section \ref{sec:SuperradiantScattering}.
We then equate the amplification factors in equations \eqref{eq:ZlmStar} and \eqref{eq:Zlm} for a massless scalar field scattering from a non-rotating star. In that limit we have $v=1$, $k=\omega$, which allows us to extract an expression for the spectral function $\rho(\omega)$ which can then be substituted into the superradiance rate \eqref{eq:SRrate} leading to
% \begin{equation}
%     \gamma_{l} = \frac{ 16 \pi R^{2\ell +3}}{\ell! (2 \ell +3)!!} \Gamma_\phi
% \end{equation}
% Substituting this into equation \eqref{eq:SRrate} we obtain the superradiance rate
\begin{align}\label{eq:UniformStar}
    &\Gamma_{n \ell m} = C_{n \ell m} \left(\frac{R}{r_{n\ell}} \right)^{(2\ell +3)}  \frac{(m \Omega- \omega )}{\omega} \Gamma_\phi,
    %   &\nonumber  {\normaltext \text{\quad \textcolor{purple}{Fran check}} }
\end{align}
where
  %  \begin{equation}
   %     C_{n \ell m} = \left(\frac{2}{n} \right)^{(2\ell +3)} \frac{1}{ n (n-l-1)! (n+l)!(2l)!! (2\ell +3)!!}.
   % \end{equation}
    \begin{equation}
        C_{n \ell m} =  \frac{ (2  \ell +1)\text{!!} (2  \ell +n+1)!}{ ( \ell +n+1) n! (2  \ell +1)!^2 (2  \ell  + 3)!!}.
        \label{eq:Cnlm}
    \end{equation}
    It is interesting to compare the amplification factor for a rotating star
    \begin{equation}\label{eq:ZlmStarRot}
Z^{\rm rot.}_{\ell m } \simeq \frac{4 R^2   (m \Omega  - \omega ) (R \omega)^{2\ell + 1}}{(1+ 2\ell )!! (2\ell + 3 )!!}\cdot\Gamma_\phi
\end{equation}
    inferred from Eq.~\eqref{eq:Zlm}, with that of a black hole
\cite{Brito:2015oca}
\begin{align}
  &Z^{\rm BH}_{lm} \nonumber \\
  &= -8GM r_+(\omega-m\Omega_{\rm H})\omega^{2\ell +1}\left(r_+-r_-\right)^{2l} \nonumber \\
  &\times \left[\frac{(\ell!)^2}{(2\ell)!(2\ell +1)!!}\right]^2\prod_{k=1}^l\left[1+\frac{(GM)^2}{k^2}\left(\frac{\omega-m\Omega_{\rm H}}{\pi r_+ T_H}\right)^2\right]
\end{align}
where $r_\pm = G M (1 \pm \sqrt{ 1-\tilde{a}^2 })$ where  $\tilde{a} = a/(GM)$ and $T_H = (r_+ - r_-)/(4 \pi r_+^2)$ is the BH temperature. Expanding this to leading order in spin and the splitting $(\omega - m \Omega_H)$, we find
\begin{align}\label{eq:ZlmBH}
  Z^{\rm BH}_{lm} &\simeq 4r_s^2 (m\Omega_{\rm H}-\omega)\left(\omega r_s \right)^{2\ell +1}  \left[\frac{(\ell!)^2}{(2l)!(2\ell +1)!!}\right]^2 \cdot \frac{1}{r_s}\
\end{align}
where $r_s = 2GM$ is the Schawrzschild radius. Hence up to combinatorial factors, we can see by comparing \eqref{eq:UniformStar} and \eqref{eq:ZlmBH} that there is a one-to-one correspondence between the superradiant scattering for stars and black holes given by replacing the Schwarzshild radius, $r_s$ with the Neutron star radius $\rightarrow R$ and identifying the damping $\Gamma_\phi$  with $1/r_s$. Hence we see that superradiance in stars will become comparable to that of stellar mass black holes when the mean free path $\lambda$, associated to the microphysics $\lambda^{-1} \simeq \Gamma_\phi$, is less than or equal to the gravitational radius of the black hole.

% {\color{red} FCD: Fixed missing factor of $\omega$. Came with a factor of 2.}

%Then you can introduce general superradiance rate (think it would be nicer also to have the general $\ell$ case here too - if you can derive it?)
%\begin{align}
%&\Delta \Gamma = \nonumber \\
%&\Gamma_{\rm em} - \Gamma_{\rm abs} \simeq  \l( \frac{GM \mu^2}{2} \r)^{5} %\frac{ \rho (m \Omega - \mu) }{2 \pi \mu} \nonumber \\
%&\simeq  \l( \frac{GM \mu^2}{2} \r)^{5} \frac{ ( m \Omega - \mu) \gamma }{2 \pi \mu} \, ,
%\end{align}

\section{Axions}\label{sec:axions}

We shall now consider the case where the bosonic field is an axion-like particle (ALP) which couples to nuclear matter within the star.  Ultra-light bosonic fields occur in many well-motivated extensions of the Standard Model. In particular, axions are an attractive dark matter candidate \cite{Kim:1979if,Shifman:1979if,Dine:1981rt,Zhitnitsky:1980tq}, as well offering a solution to the strong CP problem\footnote{Note some recent debate over whether the $\theta$ term actually appears in observables \cite{Ai:2020ptm}.} \cite{Peccei:1977hh,Weinberg:1977ma,Wilczek:1977pj} and arising generically in string compactifications \cite{Svrcek:2006yi,Arvanitaki:2009fg}. Indeed in recent years, neutron stars have become an exciting environment in which to study axions \cite{Pshirkov:2007st,Garbrecht:2018akc,Hook:2018iia,Huang:2018lxq,Battye:2019aco,Leroy:2019ghm,Darling:2020uyo,Darling:2020plz,Witte:2021arp,Battye:2021xvt,Millar:2021gzs,Battye:2021yue,Foster:2022fxn}

We shall model the  nuclear interactions using chiral perturbation theory. In this instance, the principle absorption channel is $a + N + N \rightarrow N + N$, which proceeds via the interactions \cite{Brinkmann:1988vi}
\begin{align}
\mathcal{L}_{a NN} & = \frac{g_{an}}{2m} \partial_{\mu}a\bar{N}\gamma^{\mu}\gamma_5 N,   \label{eq:aNN}\\
& \nonumber \\
 \mathcal{L}_{\pi NN} &=  \frac{2m}{m_{\pi}} f \pi_0 \bar{N}  \gamma^5 N
 \label{eq:PiNN},
\end{align}
where $f\approx 1$ and $N  = p, n$ are the proton or neutron.  

\begin{figure*}[t!]
    \centering
    \includegraphics[scale=0.8]{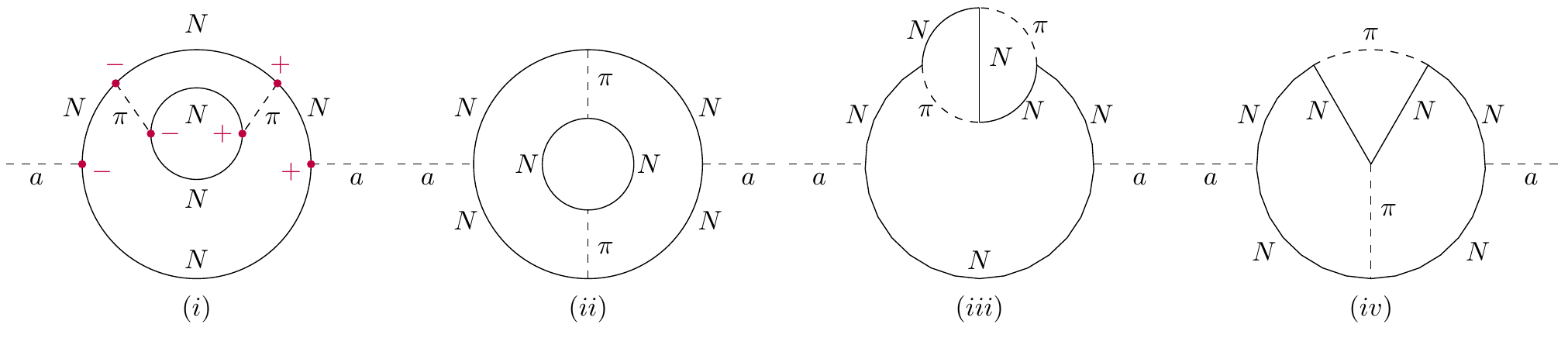}
    \caption{Axion self energies whose optical cuts give rise to $N+N+a \rightarrow N + N$ processes corresponding to ${\rm Im} \, \Pi^R$.}
    \label{fig:self energy}
\end{figure*}

Note that we have chosen a basis in which the axion-nucleon interactions are ``pseudo-vector" (PV) (axion derivatives coupling to the axial nucleon current) and the neutral pion coupling is ``pseudo-scalar" (PS) (non-derivative coupling of the pion to nucleons). Since the pion and axion are pseudo Nambu-Goldstone bosons, we can always perform chiral transformations, leading to different combinations of PV and PS couplings to nucleons. This was demonstrated explicitly in \cite{CKJ} which illustrated the various combinations of PS and PV couplings which can be achieved.  In particular, our parametrisation \eqref{eq:aNN}-\eqref{eq:PiNN} allows us to make contact with the classic reference \cite{Brinkmann:1988vi} in which axion-nucleon absorption processes are considered.

We shall now see how  the microphysical scattering processes which contribute to $ \Pi^R$ and thence to the damping term $\Gamma_a$ of the axion emerge naturally from cuts through an appropriate self-energy, in accordance with the optical theorem. For the interactions \eqref{eq:aNN}-\eqref{eq:PiNN} there are four such diagrams shown in figure \ref{fig:self energy} whose cuts lead to processes $a + N  + N \leftrightarrow N + N$. We do not consider diagrams at lower order in $\mathcal{L}_{\pi N N}$  whose cuts would lead to $a + N \leftrightarrow N + \pi$ processes. These scattering do not occur for the low energy axion modes we consider as the pion is too heavy.
 
As explained in Section \ref{sec:NeqQFT}, the spectral self-energy (which encodes information about dissipation) can be easily obtained once either of the Wightman self-energies is known (see Eq.~\eqref{eq:PiA}). Let us therefore compute $\Pi^>$, which is obtained by summing the internal vertices in the diagram over Schwinger-Keldysh indices. The internal propagators connecting different $\pm$ Schwinger-Keldysh therefore each carry $\pm$ labels. For Fermions, in momentum space, these free thermal propagators are given by (following the conventions of \cite{Prokopec:2003pj,Prokopec:2004ic})

\begin{align}\label{eq:FermionPropagators}
i {\cal S}^{++} (p)  &= \phantom{-} \frac{i (\slashed{p}+m)}{p^2 - m^2 + {\rm sign}(p^0) \epsilon} \nonumber \\
&- 2 \pi (\slashed{p}+m)\delta(p^2-m^2){\rm sign}(k^0)  \, F(p^0) ,\nonumber \\
i {\cal S}^{--}(p) &= - \frac{i (\slashed{p}+m)}{p^2 - m^2 + i{\rm sign}(p^0) \epsilon} \nonumber \\
& + 2 \pi (\slashed{p}+m){\rm sign}(p^0) \delta_D(p^2-m^2) \, \left(1 -F(p^0) \right) \nonumber \\
i {\cal S}^{+-}(p)  &= - 2\pi \, {\rm sgn}(p^0) \, (\slashed{p}+m)\, \delta_D(p^2-m^2) \,  F(p^0), \nonumber \\
i {\cal S}^{-+}(p) &=  2\pi \, {\rm sgn}(p^0) \, (\slashed{p}+m)\, \delta_D(p^2-m^2)  \left[1-F(p^0) \right],  \nonumber \\
\end{align}
% \textcolor{purple}{((2.69)-(2.72) of hep-ph/0312110.}
For bosons the relevant propagators are

\begin{align}\label{eq:BosonPropagators}
i {\cal D}^{++} (p) & = \frac{i}{p^2 - m^2+ i{\rm sign}(p^0) \epsilon} \nonumber \\ 
&+ 2 \pi \delta_D(p^2-m^2) \,{\rm sign}(p^0) B(p^0),  \nonumber \\
i {\cal D}^{--} (p) & = -\frac{i}{p^2 - m^2+ i{\rm sign}(p^0) \epsilon} \nonumber \\
&+ 2 \pi \delta_D(p^2-m^2) \,{\rm sign} (p^0) (1+B(p^0)),  \nonumber \\
\end{align}
% \textcolor{purple}{((2.64) of hep-ph/0312110.}
Here
\begin{equation}\label{eq:FD_BE}
    F(p^0) = \frac{1}{e^{(p^0 - \mu)/T} + 1} ,\quad B(p^0) = \frac{1}{e^{(p^0 - \mu)/T} - 1}
\end{equation}
are the Fermi-Dirac and Bose-Einstein distributions respectively with chemical potential $\mu$.

Let us start by computing diagram (i) of Fig.~\ref{fig:self energy}. As stated above, we are only interested in processes $a+ N+ N \leftrightarrow  N + N$ which correspond to the explicit choice of internal Schwinger Keldysh indices shown. The other choices of internal indices would correspond to cuts resulting in processes $a \leftrightarrow N + N$ or $a+ N \leftrightarrow N + \pi$ which are kinematically forbidden for the low-energy axions considered here. The only choice of indices giving rise to $a + N+ N \leftrightarrow N+ N$ processes corresponds to a cut down the centre line. The corresponding expression for this contribution to $\Pi^>$ is then given by   
%\begin{widetext}
%\begin{equation}
%\begin{aligned}
 %   \Pi^{>} = -i G_{\pi n}^4 G_{a n}^2 \int & \frac{d^4 k}{(2 \pi)^4} \frac{d^4 r}{(2 \pi)^4} \frac{d^4 p}{(2 \pi)^4} {\rm Tr} % \left[\gamma^5 \slashed{q} S^{+-}(k) \gamma^5 \slashed{p} S^{--}(q+k) \gamma^5 \slashed{p} S^{-+}(q+k+p) \gamma^5 \slashed{q} % S^{--}(q+k)\right] \\
 %   & {\rm Tr} \left[S^{+-}(r) \gamma^5 \slashed{p} S^{-+}(r-p) \gamma^5 \slashed{p}\right] D^{--}(p) D^{++}(p),
%\end{aligned}
%\end{equation}
% \end{widetext}

\begin{widetext}
% \begin{equation}
% \begin{aligned}\label{eq:Pi_GeneralExpr}
%  \Pi^{>}_{(i)}(q) = -i  G_{\pi n}^4 G_{a n}^2 \int & \frac{d^4 k}{(2 \pi)^4} \frac{d^4 r}{(2 \pi)^4} \frac{d^4 p}{(2 \pi)^4}
%   {\rm Tr}  \left[\gamma^5 \slashed{q} S^{+-}(k) \gamma^5 \slashed{q} S^{--}(q+k) \gamma^5 S^{-+}(q+k-p) \gamma^5 S^{++}(q+k)\right] \\
%     & {\rm Tr} \left[\gamma^5 S^{+-}(r) \gamma^5 S^{-+}(r+p) \right] D^{--}(p) D^{++}(p), \textcolor{red}{\text{Change to $p_{i=1-4}$.}}
% \end{aligned}
% \end{equation}
% \begin{equation}
\begin{align}\label{eq:Pi_GeneralExpr}
 &i\,\Pi^{>}_{(i)}(q)  =   \nonumber \\ 
 &\left(-\frac{i g_{an}}{2m}\right)^2  \left(-  \frac{i 2 m f }{m_{\pi}} \right)^4 \int  \frac{d^4 p_1}{(2 \pi)^4} \frac{d^4 p_2}{(2 \pi)^4} \frac{d^4 p_3}{(2 \pi)^4}\frac{d^4 p_4}{(2 \pi)^4}
 (2 \pi)^4 \delta^{(4)}(q+p_1+p_2 - p_3 - p_4)  i {\cal D}^{--}(p_4 - p_2) i {\cal D}^{++}(p_4 - p_2), \nonumber \\
 &
 {\rm Tr}  \left[\gamma^5 i\slashed{q} iS^{+-}(p_1) \gamma^5 i \slashed{q} i {\cal S}^{--}(q+p_1) \gamma^5 i {\cal S}^{-+}(p_3) \gamma^5 i {\cal S}^{++}(q+p_1)\right]
     {\rm Tr} \left[\gamma^5 i {\cal S}^{+-}(p_2) \gamma^5 i {\cal S}^{-+}(p_4) \right],    
\end{align}
% \textcolor{purple}{I followed the same $i$ conventions as when computing the 1-loop self-energy in Eq. (2.8) of hep-ph/0406140. i get two factors of $(-1)$ from the Fermion loops (Grassmann variables) which cancel. I keep the i prefactors in front of the propagators functions. The $i$'s from the pion couplings and propagators occur in multiples of 4, so have no effect. The only one to worry about is that arising from the axion-coupling which can change the overall sign, and also the two $i$'s which arrive implicitly from the axion derivative coupling, which again would change the overall sign. The factors of $(2\pi) $ also seem consistent with (2.8) of Prokopek II. }% We now define a new set of momentum variables $p_1 = k$, $p_2 = r$, $p_3 = q+k-p$, $p_4 = p+r$. In terms of these variables
%  \textcolor{red}{JMcD: can you (Fran) change the form of \eqref{eq:Pi_GeneralExpr} so it looks like a 4 particle phase space integral (i.e. don't use the overall delta function) }
where $q$ is the axion momentum. We can now insert the explicit expressions \eqref{eq:FermionPropagators}-\eqref{eq:BosonPropagators} for the propagators. The Feynman and anti-Feynman propagators with indices $++$ and $--$, respectively, have a vacuum (virtual) and thermal on-shell part. The finite temperature piece can again be neglected on the grounds it corresponds to thermal corrections to the pion and electron propagators, which are again suppressed inside the neutron star since $T/m_\pi,\, T/m \ll 1$.
% \textcolor{red}{JMcD: Bj\"orn, is this the correct argument, or is that they are on shell and so somtimes corresopnd to a kinematically forbidden vertex? }.
As a result, these propagators are dominated by their vacuum zero-temperature piece, corresponding to the usual propagation of virtual particles. This allows us to make connection with zero-temperature matrix elements. With these approximations in mind, after taking the Dirac traces, the remaining integral takes the form
\begin{align}\label{eq:Pi_MatrixElt}
   i\,\Pi^{>}_{(i)}(q) = & \int  \frac{d^4 p_1}{(2 \pi)^3} \frac{d^4 p_2}{(2 \pi)^3} \frac{d^4 p_3}{(2 \pi)^3} \frac{d^4 p_4}{(2 \pi)^3} (2 \pi)^4 \delta^{(4)}(q+p_1+p_2 - p_3 - p_4) \nonumber \\
  &{\rm sgn}\,(p_1^0)  \delta(p_1^2 - m^2)  {\rm sgn}\,(p_2^0)  \delta(p_2^2 - m^2)  {\rm sgn}\,(p_3^0)  \delta(p_3^2 - m^2){\rm sgn}\,(p_4^0)  \delta(p_4^2 - m^2) \nonumber \\
  &F(p_1^0)F(p_2^0)  (1-F(p_3^0)) (1-F(p_4^0))   \, C (q,p_1,p_2,p_3,p_4),
\end{align}
where $C$ is an effective matrix element which results from the trace parts (to leading order in axion momentum $q_\mu$)
\begin{align}\label{eq:Cdef}
   & C(q,p_1,p_2,p_3,p_4)   \nonumber \\
   & = \frac{4 g_{a n}^2 f^4 m^2} {m_\pi^4}  \frac{1}{((p_4 - p_2)^2 - m_{\pi}^2)^2}\frac{16 \left[ 4 (q \cdot p_1)^2 (m^2 - p_1 \cdot p_3)+  q^2 m^2( 6 p_1\cdot p_3  - 4 m^2 ) \right](m^2 - p_2 \cdot p_4)}{((q+p_1)^2 - m^2)^2},
\end{align}
In what follows we will drop terms directly involving $q^2$ but retain those arising from $(p_i\cdot q)^2$ in order to make contact with the estimates made in e.g. \cite{Harris:2020rus,Brinkmann:1988vi}. In those works, the authors had in mind relativistic (i.e. lightlike) axions, and so they did not retain terms $q^2$ in the matrix elements. In the traces, this is equivalent to keeping terms such as $\textbf{p}_i.\textbf{p}_j (p.q)^2 \sim p_F^2 m^2 m_a^2$ whilst dropping terms $p^4 q^2 \sim m^4 m_a^2$. We repeat this step, allowing us to quote existing results from the literature involving matrix elements and their interferences. For the non-relativistic axions considered here, the latter terms would in fact enhance the matrix element by $~ (m/p_F)^2 $ so this approximation is actually conservative. We leave these corrections for future work, with the view that our main goal is to demonstrate a general procedure for computing superradiant rates.  

Next we must bring Eq.~\eqref{eq:Pi_MatrixElt} into the form of a phase-space integral over on-shell modes. This we do by expanding the delta functions using
$\delta(p^2 - m^2) = \frac{\delta(p^0 - E_p)}{2 E_p} + \frac{\delta(p^0+E_p)}{2 E_p}$ where $E_p = \sqrt{\textbf{p}^2 + m^2}$. We also use the relation $F(-E) = 1-{\bar F}(E)$, where ${\bar F}(E)$ is the phase space distribution for the antiparticle given by replacing $\mu \rightarrow - \mu$ in Eq.~\eqref{eq:FD_BE}. Using these relations and performing the $p^0$ integrations we arrive at
\begin{equation}
\begin{aligned}\label{eq:self-energyFUll}
    i \Pi^{>}_{(i)}(q) = \int & \frac{d^3 p_1}{2 E_1(2 \pi)^3} \frac{d^3 p_2}{2 E_2 (2 \pi)^3} \frac{d^3 p_3}{2 E_3 (2 \pi)^3} \frac{d^3 p_4}{2 E_4 (2 \pi)^3}  (2 \pi)^4 \delta^{(3)}({\bf q}+{\bf p_1}+ {\bf p_2} - {\bf p_3} - {\bf p_4}) \\ &  [C(q,p_1,p_2,p_3,p_4)F(E_1)F(E_2)(1-F(E_3))(1-F(E_4)) \delta(E_q+E_1+E_2 - E_3 - E_4)  + \\& C(q,p_1,\bar{p}_2, \bar{p}_3,p_4)F(E_1){\bar F}(E_3)(1-{\bar F}(E_2))(1-F(E_4)) \delta(E_q+E_1+E_3 - E_2 - E_4)   + \\& C(q,\bar{p}_1,p_2,\bar{p}_3,p_4) {\bar F}(E_3)F(E_2)(1-{\bar F}(E_1))(1-F(E_4)) \delta(E_q+E_3+E_2 - E_1 - E_4) + \\& C(q,p_1,\bar{p}_2,p_3,\bar{p}_4) F(E_1){\bar F}(E_4)(1-F(E_2))(1-{\bar F}(E_3)) \delta(E_q+E_1+E_4 - E_2 - E_3)  + \\& C(q,\bar{p}_1,p_2,p_3,\bar{p}_4) F(E_2){\bar F}(E_4)(1-{\bar F}(E_1))(1-F(E_3)) \delta(E_q+E_2+E_4 - E_1 - E_3)  + \\& C(q,\bar{p}_1,\bar{p}_2,\bar{p}_3,\bar{p}_4){\bar F}(E_3){\bar F}(E_4)(1-{\bar F}(E_1))(1-{\bar F}(E_2)) \delta(E_q+E_3+E_4 - E_1 - E_2)],
\end{aligned}
\end{equation}
where we have used the fact that the energy conserving delta function can only be satisfied if there are two neutrons each in the initial and final states and $p, \bar{p} = (\pm E_p, {\bf p})$  gives the on-shell 4-momenta of positive and negative energies. Note we have also neglected the momentum transfer associated to the axion in the $3$-momentum delta function.

% Cite Kim paper on pion/axion reparametrisation of PV vs. PS interactions:
% \url{https://journals.aps.org/prl/pdf/10.1103/PhysRevLett.62.849}
%By relabelling the integration variables and taking e.g. ${\bf p} \rightarrow {\bf -p}$ as needed to preserve the structure of the momentum conserving delta function, we can recast this integral into a more familiar form:
%\begin{widetext}
%\begin{equation}
%\begin{aligned}
 %   \Pi^{>} = -i G_{\pi n}^4 G_{a n}^2 \int & \frac{d^3 p_1}{2 E_1(2 \pi)^3} \frac{d^3 p_2}{2 E_2 (2 \pi)^3} \frac{d^3 p_3}{2 E_3 (2 \pi)^3} \frac{d^3 p_4}{2 E_4 (2 \pi)^3}  (2 \pi)^4 \delta^{(4)}(q+ p_1+ p_2 - p_3 - p_4) f(E_1)f(E_2)(1-f(E_3))(1-f(E_4)) \\ &  [C(q,p_1,p_2,p_3,p_4)  + C(q,p_1,-p_3,-p_2,p_4) +  C(q,-p_3,p_2,-p_1,p_4) + \\& C(q,p_1,-p_4,p_3,-p_2) + C(q,-p_4,p_2,p_3,-p_1)  +  C(q,-p_3,-p_4,-p_1,-p_2)],
%\end{aligned}
%\end{equation}
%\end{widetext}
 Next we assume that we are in the regime $\mu /T \gg 1$, so that neutrons dominate over anti-neutrons. Under these approximations, all those terms containing a $\bar{F}$ in the initial state can be discarded, leading to
\begin{align}\label{eq:SimpleCollIntegral}
   i \Pi^{>}_{(i)}(q) =   \int & \frac{d^3 p_1}{2 E_1(2 \pi)^3} \frac{d^3 p_2}{2 E_2 (2 \pi)^3} \frac{d^3 p_3}{2 E_3 (2 \pi)^3} \frac{d^3 p_4}{2 E_4 (2 \pi)^3}   \nonumber \\
    & (2 \pi)^4\delta^{(4)}(q+ p_1+ p_2 - p_3 - p_4) F(E_1)F(E_2)(1-F(E_3))(1-F(E_4))  C_{(i)}(q,p_1,p_2,p_3,p_4). \\
\end{align}
\end{widetext}
% where we have dropped the indices on $p^+$ so that the $p_i$ should be read as $p_i = (E_{\textbf{p}_i},\textbf{p}_i)$.
Thus we have brought the self-energy into the form of a standard collision integral. This again represents the power of thermal and non-equilibrium field theory not only to derive an effective equation of motion for the scalar field, but also to express the damping coefficients in that equation in terms of straightforward collision integrals. 

\begin{figure}[t!]
    \centering
    \includegraphics[scale=0.55]{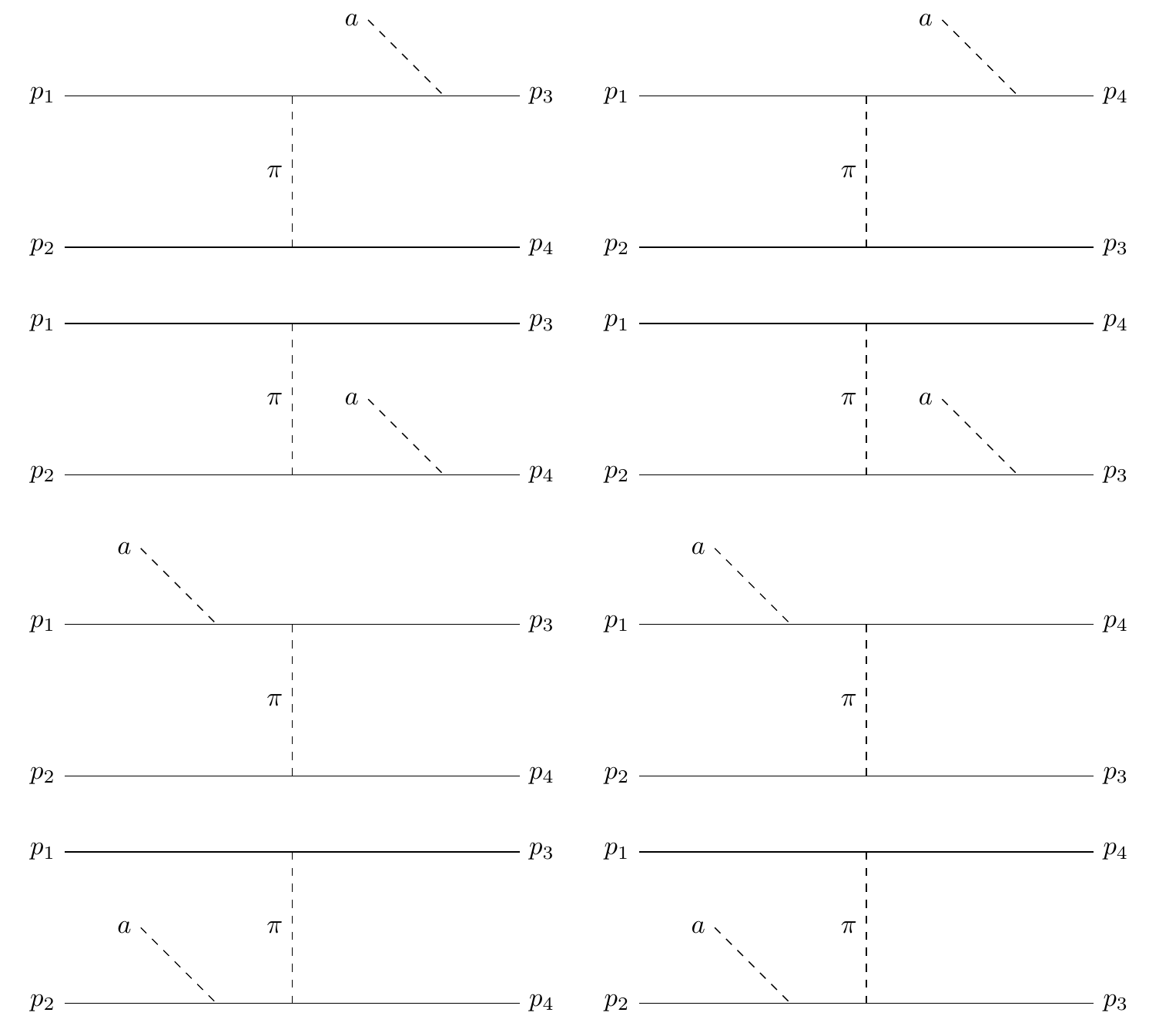}
    \caption{Matrix Elements for axion absorption corresponding to the cuts through neutron lines in Fig.~\ref{fig:self energy}. The left and right columns correspond to $t$- and $u$-channel processes respectively. }
    \label{fig:MatrixElts}
\end{figure}

To make contact with the calculations in \cite{Brinkmann:1988vi} we now go to the non-relativistic limit and expand the matrix element to leading order in $\textbf{p}_i/m$.
% where the four-momenta now take their standard positive energy values.
In the limit of non-relativistic neutrons, we find, to leading order in $q$:
% \begin{equation}
%     C_{(i)}(q,p_1,p_2,p_3,p_4) = G_{\pi n}^4 G_{a n}^2 \frac{4 ({\bf p_1} \cdot {\bf p_3}) ({\bf p_2} \cdot {\bf p_4})}{(|{\bf p_4} - {\bf p_2}|^2+m_{\pi}^2)^2}
% \end{equation}
% Using the relations $\textbf{p}$
% Now let us assume that all neutron momenta take values ${\bf p} \sim T$ and that the interaction is a glancing collision such that ${\bf p_1} \cdot {\bf p_3} \sim {\bf p_2} \cdot {\bf p_4}\sim T^2{\rm cos}(\theta)$ with $\theta = \pi + \delta \theta$ and $\delta \theta \ll 1$. Then we have ${\bf p_1} \cdot {\bf p_3} \sim {\bf p_2} \sim \frac{|{\bf k}|^4}{16}$.
\begin{equation}
     C(q,p_1,p_2,p_3,p_4) \simeq   \frac{32 f^4 g_{a n}^2 m^2}{m_{\pi}^4}\frac{|{\bf k}|^4}{(|{\bf k}|^2+m_{\pi}^2)^2}.
\end{equation}
where $\textbf{k} = \textbf{p}_2 - \textbf{p}_4$. Note this is precisely the form of the matrix element associated to the modulus-squared of the top-left diagram in Fig. \ref{fig:MatrixElts}, as can be seen from the appendix of ref. \cite{Brinkmann:1988vi}.
% where we used  $G_{\pi n} = \frac{2 m g_{\pi n}}{m_\pi}$ and $G_{a n} = \frac{g_{a n}}{2 m}$.
% To leading order in axion momentum and in the same non-relativistic limit, one has
% \begin{equation}
%  \sum_{\text{spins}}  \left| \mathcal{M}_{(a)} \right|^2 = \frac{16 f^4 g_{\pi n}^4 g_{a n}^2 m^2}{m_{\pi}^4}\frac{|{\bf k}|^4}{(|{\bf k}|^2+m_{\pi}^2)^2}.
% \end{equation}
Thus, as expected by the optical theorem, cuts through diagrams in fig.~\ref{fig:self energy} correspond to $\Pi^>$ which feeds into the imaginary part of the retarded self-energy. In turn these cuts can be interpreted in terms of S-matrix elements for the process $a+ N + N \rightarrow N + N$.

% From a book-keeping perspective correspondence between matrix elements and cuts through diagrams is not a simple one-to-one: for example, the matrix element $(a)$ receives contributions from cuts through both self-energy diagrams $(i)$ and $(ii)$. Nonetheless, as demanded by the optical theorem and within the approximations made here (where we neglect thermal corrections to the pion and neutron Feynman/anti-Feynman propagators), at the level of the full collision integral, the sum over all diagrams $(i)$-$(iv)$ must produce the same collision integral as naively integrating the square of all matrix elements and their interferences.  

One can repeat this process for other diagrams in Fig.~\ref{fig:self energy} which can be similarly identified with matrix elements (and their interferences) in Fig.~\ref{fig:MatrixElts}. This is of course just the optical theorem, which demands that the sum over diagrams $(i)$-$(iv)$ must give
\begin{widetext}
\begin{align}\label{eq:SimpleCollIntegral2}
    i \Pi^{>}(q) =  \int & \frac{d^3 p_1}{2 E_1(2 \pi)^3} \frac{d^3 p_2}{2 E_2 (2 \pi)^3} \frac{d^3 p_3}{2 E_3 (2 \pi)^3} \frac{d^3 p_4}{2 E_4 (2 \pi)^3}   \nonumber \\
    & (2 \pi)^4\delta^{(4)}(q+ p_1+ p_2 - p_3 - p_4) F(E_1)F(E_2)(1-F(E_3))(1-F(E_4))   \sum_{\text{spins}} \vert \mathcal{M}\vert^2
\end{align}
where $\mathcal{M}$ is the full matrix element associated to the diagrams in Fig.~\ref{fig:MatrixElts}.  This is a classic result which has already been computed a long time ago \cite{Brinkmann:1988vi,Iwamoto:1984ir}
\begin{align}
\label{eq:FullMSqd}
\sum_{\text{spins}} \vert \mathcal{M}\vert^2  = \frac{256}{3}\frac{f^4m^2 g_{an}^2}{m_{\pi}^4} \left[ \frac{\mathbf{k}^4}{\left(\mathbf{k}^2+m_{\pi}^2\right)^2}+\frac{\mathbf{l}^4}{\left(\mathbf{l}^2+m_{\pi}^2\right)^2}+\frac{\mathbf{k}^2\mathbf{l}^2-3\left(\mathbf{k}\cdot\mathbf{l}\right)^2}{\left(\mathbf{k}^2+m_{\pi}^2\right)\left(\mathbf{l}^2+m_{\pi}^2\right)}    \right].
\end{align}
\end{widetext}
Note that strictly speaking, as can be seen in \cite{Brinkmann:1988vi}, this is actually the matrix element given by averaging over the direction of the axion three-momentum, so that in the matrix element, one replaces e.g. $(\hat{ \textbf{k} } \cdot \hat{\textbf{q} } )^2 \rightarrow \braket{(\hat{ \textbf{k} } \cdot \hat{\textbf{q} } )^2 } =1/3$. Here hats denote unit vectors. However, by symmetry we know the final self energy $\Pi^>(q) =\Pi^>(q^0,\textbf{q}) $ \textit{must} be isotropic with respect to $\textbf{q}$ so that $\Pi^>(q^0,\textbf{q}) =\Pi^>(q^0, \left| \textbf{q} \right| )$. It therefore follows that the self-energy is invariant under the averaging over the direction of $\hat{\textbf{q}}$.

Formally this means $\Pi^> = \braket{\Pi^>}$. Furthermore, when applying the operator $\braket{\cdots}$ to Eq.~\eqref{eq:SimpleCollIntegral}, the averaging can be taken past the distribution functions and energy conserving delta functions (provided we neglect the axion 3-momentum in the spatial 3-momentum conserving delta function) and directly onto the matrix element in the collision integral. As a result, to leading order in axion-momentum, the answer is unchanged by using a matrix element averaged over the direction of axion momentum.  

Having related the self-to a collision integral \eqref{eq:SimpleCollIntegral2}, whose matrix element \eqref{eq:FullMSqd} is known, all that remains is to actually compute this integral. Fortunately, this integral can be directly identified with the so-called axion mean free path $\lambda^{-1}_>$ defined by \cite{Ishizuka:1989ts,Burrows:1990pk,Iwamoto:1992jp,Dietrich:2019shr,Harris:2020rus}
\begin{equation}\label{eq:MFP}
   \lambda^{-1}_> = \frac{i \Pi^>(q)}{2 q^0}.
\end{equation}
The identification of a $>$ subscript with the mean free path will become clearer shortly. For now though, we simply note that by looking up the pre-existing calculations of the mean-free path $\lambda^{-1}_>$ in the literature, we can immediately read off $\Pi^>$. The damping rate then follows immediately from Eq.~\eqref{eq:PiA}.

In the degenerate limit appropriate for neutron stars, this integral has in fact already been computed in a Fermi Surface approximation
\cite{Ishizuka:1989ts,Iwamoto:1992jp,Harris:2020rus}, in that limit one finds
\begin{equation}
\lambda^{-1}_> = \frac{1}{18\pi^5}\frac{f^4g_{an}^2m^2}{4 m_{\pi}^4}p_{Fn}Q(y)\frac{\omega^2+4\pi^2T^2}{1-e^{-\omega/T}},\label{eq:lambdaFS}
\end{equation}
where
\begin{align}
Q(y) =& 4 - \frac{1}{1+y^2} + \frac{2y^2}{\sqrt{1+2y^2}}\arctan{\left(\frac{1}{\sqrt{1+2y^2}}\right)} \nonumber \\
&-5y\arcsin{\left(\frac{1}{\sqrt{1+y^2}}\right)},
\label{eq:Fofy}
\end{align}
and $y= m_\pi/(2 p_F)$ with $p_F$ the neutron Fermi momentum. We can combine Eq.~\eqref{eq:PiA} with the KMS relation $\Pi^>(q) = e^{\omega/T}\Pi^<(q)$ to obtain
\begin{align}
   {\rm Im}[\Pi^R(p,x)]& =    - \frac{i}{2}\left(\Pi^>(p,x)-\Pi^<(p,x)\right) \nonumber \\
   &= - \frac{i}{2}\left(1- e^{-\omega/T}\right)\Pi^>(p,x) .
\end{align}
Inserting \eqref{eq:MFP} and \eqref{eq:lambdaFS} into this we can immediately compute the anti-hermitian part of the self energy in the degenerate limit, which gives
\begin{equation}\label{eq:DampingNucleon}
    \text{Im} \, \Pi^R =  \textcolor{purple}{-}\frac{\omega}{18\pi^5}\frac{f^4g_{an}^2m^2}{4 m_{\pi}^4}p_{F}Q(y) \left( \omega^2+4\pi^2T^2 \right)
\end{equation}

Putting this together, from Eq.~\eqref{eq:SupRatePi}, we are able to get a superradiance expressed in terms of microphysical parameters:
%\begin{widetext}
%\begin{equation}
 %   \Gamma^{a}_{n \ell m} =  \frac{1}{2}  \frac{1}{18\pi^5}\frac{f^4g_{an}^2m^4}{m_{\pi}^4}p_{F}F(y) 4\pi^2T^2  \left( \frac{2 R}{r_{n \ell}} \right)^{2\ell +3} \frac{l!(2\ell +1)!!}{8 \pi n (n-l-1)! (n+l)! (2\ell +1)!}  (m \Omega - \omega ).
%\end{equation}
%\end{widetext}

\begin{widetext}
\begin{equation}
    \Gamma^{a}_{n \ell m} =  \frac{1}{18\pi^5}\frac{f^4 g_{an}^2m^2}{4 m_{\pi}^4}p_{F}Q(y) 4\pi^2T^2  \left( \frac{R}{r_{n \ell}} \right)^{2\ell +3} \frac{ (2 \ell+1)\text{!!} (2 \ell+n+1)!}{ (\ell+n+1) n! (2 \ell+1)!^2 (2 \ell + 3)!!} \frac{(m \Omega - \omega_{\ell n} )}{\omega_{\ell n}},
 \end{equation}
\end{widetext}
 Let us make some estimates of this rate. We can take typical neutron star values $\Omega = 600 {\rm Hz}$, $M= 1.4 M_\odot$ and $R=10{\rm km}$ and $p_F \simeq 300 {\rm MeV}$. For couplings constants we can take $f\simeq 1$ and the maximum limit set by Supernova constraints is $g_{a n} \simeq 10^{-9}$. Furthermore, for hot young stars, the rate is at its highest, here (although briefly) temperatures could reach up to $T  = 10^{11}{\rm K}$ \cite{Raffelt:1996wa} giving an upper bound on the superradiance rate. For these numbers, the $\ell =m=1, n=0$ has a superradiance timescale $\Gamma_{0 1 1}^{-1}\simeq 10^{13}{\rm yr}$, somewhat longer than the age of the Universe. For older, cooler neutron stars, the rate would of course be lower. Hence at least for the simplest axion scenarios which couple axions to nuclear matter, the star is protected from superradiant instabilities over timescales of the Universe.

At this point a few remarks are in order concerning the microphysical damping rate as embodied in our Eq.~\eqref{eq:DampingNucleon} and the notion of the mean-free path found elsewhere in the literature \cite{Ishizuka:1989ts,Burrows:1990pk,Iwamoto:1992jp,Dietrich:2019shr,Harris:2020rus}. Formally the damping rate of the field in the equation of motion is defined as the difference of two Wightman self energies
\begin{equation}
    \text{Im} \, \Pi^R  =
    \frac{1}{2i}\left(\Pi^> -  \Pi^<  \right)   
\end{equation}
where, by virtue of the KMS relation, we can write the self-energies more transparently as
\begin{equation}
    i \Pi^> = - (1+B) \text{Im} \Pi^R, \qquad  i \Pi^< = - B \,\text{Im} \Pi^R,
\end{equation}
where $B$ is the Bose-Einstein distribution function defined in \eqref{eq:FD_BE}. Notice therefore that the \textit{true} mean free path involves two \textit{opposite} contributions where $\Pi^>$ and $\Pi^<$ are to be interpreted as gain and loss processes, respectively, as is apparent from the presence of the $B$ and $1+B$ prefactors.  Notice however, that the collision integral in  \cite{Harris:2020qim} which the authors claim corresponds to the mean free path, actually only give the piece  $\Pi^>$, explaining the appearance of a prefactor $1/(1-e^{-\omega/T}) = 1 + B$. In reality, the authors should have taken both contributions, which would have given a mean free path $\lambda^{-1} = (\text{Im} \Pi^R/\omega)$.  

The fact that it is ${\rm Im} \Pi $ and not just one of $\Pi^>$ or $\Pi^<$ which should be identified with the mean free path is also apparent if one write the spectral propagator of the axion $G_s(q)$ in Breit-Wigner Form \cite{Blaizot:2001nr}
\begin{equation}
    G_s(q) = \frac{ 2\text{Im} \Pi^R}{ \left( (q^0)^2 - E_\textbf{q} - \text{Re}\,\Pi^R\right)^2 +  (\text{Im} \Pi^R)^2}.
\end{equation}
This spectral propagator describes the propagating axion modes in the medium. In the lossless limit $\text{Im} \Pi^R \rightarrow 0$, we recognise the above form as the Lorenz representation of the delta function so that the spectral function $G_s(q)$ simply becomes a mass-shell condition $G_a(q) \propto \delta\left((q^0)^2 - E_\textbf{q} - \text{Re}\,\Pi^R\right)$ of a zero width mode where $\text{Re}\,\Pi^R$ encodes the so-called dispersive part of $\Pi^R$ which gives radiative corrections to the axion dispersion relation. This provides yet another justification that $\text{Im} \, \Pi^R$  sets the width of axion states associated to the lifetime of the axion in-medium.

The question is then whether the authors of \cite{Harris:2020qim} misidentifying the mean free path with the wrong type of self-energy has any quantitative bearing on their calculation. Since there the authors typically consider axions with energies $\omega \gtrsim T$, we have that $B \lesssim 1$, so that proved one considers axion frequencies at or above the temperature, $B$ factors become negligible and one finds $\text{Im} \Pi^R \simeq \Pi^>/2i$. This misidentification would therefore only change the mean free path by $\mathcal{O} (1)$ so long as one sticks to $\omega \gtrsim T$.  

In fact, the contribution to the damping rate associated to $\Pi^>$ corresponds to stimulated emission of axions. This means that for superradiance, since we are in the regime $\omega \ll T$ we have $B>>1$ owing to large bose enhancement. This means that the two Wightman contributions almost cancel, with their difference being suppressed by $\omega/T$ so that to leading order in this ratio we have $\text{Im} \Pi^R \simeq \frac{\omega}{T} \Pi^> $. Hence in our case, since $\omega/T \simeq 10^{-17}$ for $T=10^8$K and $\omega = 10^{-13}$eV, this would lead to a radically different estimate for the superradiance rate if we had misidentified the wrong collision integral (associated to $\Pi^>$) with the true damping rate of the field.  This again emphasises the importance of carrying out a rigorous first principles calculation of the damping rate within non-equilibrium field theory and not simply guessing it from a collision integral.

% Where $\Pi^>$. One can see from the presence of the $1+B$ corresponds to the

% \textcolor{red}{Be careful to illustrate the difference between the spectral self energy (microphysics) and spectral propagator/correlator (EFT). }

% At this stage, we should perhaps return

% Djuna's paper on energy loss in supernovae \cite{Sakstein:2022tby} also

% EXPLAIN WHAT IS DRIVING THE SUPRESSION - IS IT DEGENERACY, OR THE AXION COUPLING. IF WE IDENTIFY $\gamma \leftrightarrow 1/r_g$ then WHY IS GRAV SUPERRADIANCE NOT PLANCK SUPPRESSED?

\section{Discussion and Conclusions}\label{sec:conclusions}

In this paper we have presented the first general treatment of superradiance in stars, which through straightforward adaptation allows the reader to compute the superradiant instability for any fundamental interaction.

One key development was to employ the full power of field theory at finite density. As shown in Section \ref{sec:NeqQFT}, this allowed us to understand how the damping of fields should be identified with the self energy as embodied in Eq.~\eqref{eq:DampedScalar} and what precise approximations are necessary to produce a simple damping term. Crucially, we were then able to relate the self-energy (and thence this damping coefficient) to microphysical scattering processes, as illustrated in the case of axions in Section \ref{sec:axions}.

By contrast, in previous work a damping term was added in a somewhat schematic way \cite{Cardoso:2015zqa} and it remained opaque as to how to relate this to microphysics. It is worth noting that \cite{Cardoso:2017kgn} made some attempt to interpret the damping coefficients for gauge field in terms of QED-like processes but it remained unclear how these damping terms emerge in a more general theories and indeed how they can be justified from a first principles approach in QFT.

% From a technical perspective, we saw how a truncation of the gradient expansion appearing in \eqref{eq:PiDiamond} was necessary in order to render the equations of motion for the field in terms of a simple damping term.
% We recognised this approximation as consistent with the neglect of spatial dispersion, as would be encountered in electromagnetism when one approximates the permittivities $\varepsilon(\omega, \textbf{k}) \rightarrow \varepsilon(\omega, 0)$ as being independent of wavelength. This allows one to go from an integro-differential equation in the response function, to a simpler differential equation for the damping of fields in terms of a \textit{local} damping coefficient.

We are now able to justify and explain the existence of damping terms for fields, and explicitly calculate them in terms of microphysical absorption processes. In the case of axions, this corresponded to inverse Bremsstrahlung processes $a + N + N \rightarrow N + N$. The final expression for the damping coefficient takes the form of a collision integral, which for the present interactions had already been derived elsewhere in the literature when computing the mean free path of axions in the context of energy transport in neutron stars.

Having computed the microphysical damping coefficient for the scalar field, it was then straightforward to compute the absorption rate of long-wavelength modes into the whole star. Our next key insight was that this was sufficient to compute the superradiant instability rates for massive fields in gravitational bound states of a \textit{rotating} star. This we achieved through appealing to an EFT of long-wavelength modes \cite{Endlich:2016jgc}. This approach allowed us to compute the absorption of arbitrary multiple modes for a static, non-rotating star, and then through symmetry arguments, relate this through a simple algebraic substitution for the rate of a rotating star. This completely circumvents the need to introduce bulk currents for the rotating matter in the star when computing the  effective equation of motion \eqref{eq:DampedScalar} as was done in \cite{Cardoso:2017kgn}. Thus rather elegantly, one simply needs to compute the absorption of a massless mode into a non-rotating star, and the superradiance rate can be derived in a quick and straightforward way.

In summary, we now have a fully working pipeline to compute superradiance rates in stars for any interaction. Having presented this general treatment, one can now of course generalise this to a broader class of couplings and fields to see for which fundamental interactions the rate is highest, and investigate their corresponding signatures.

\begin{center}
\textbf{Acknowledgements}
\end{center}
JIM is supported by an FSR Incoming Postdoctoral Fellowship. FCD is supported by Stephen Hawking Fellowship EP/T01668X/1 and STFC grant ST/T001011/1. BG acknowledges
support from the Collaborative Research Centre SFB 1258 of the
Deutsche Forschungsgemeinschaft. We thank M.~Drewes for useful conversations. We are indebted to Paolo Pani, Richard Britto and Vitor Cardoso for comments on the manuscript and to Steven Harris for his input on axion mean free paths. 

\begin{center}
\textbf{Data Access Statement}
\end{center}
No data sets are used or created in this paper.

\appendix

\section{Convolutions of Green Functions}\label{App:Convolution}
The full derivation of the result \eqref{convolution:homogeneous} is as follows. We begin with
 \begin{align}
        &[\Pi^R\odot\phi](x)=\int d^4 y \Pi^R(x,y)\phi(y)\notag\\
        =&\int d^4 y \int\frac{d^4 p}{(2\pi)^4}\Pi^R\left(p,\frac{x+y}{2}\right)\phi(y) e^{-ip\cdot(x-y)}.
\end{align}
where we defined a convolution operator $\odot$ to compactify notation. Expanding $\Pi^R$ about $p=0$ to all orders, we obtain
\begin{align}
       &[\Pi^R\odot\phi](x) \nonumber \\ =&\sum_{n=0}^\infty \int d^4 y \int\frac{d^4 p}{(2\pi)^4} \frac{p^n}{n!} \left[\partial^n_p \Pi^R\left(p,\frac{x+y}{2}\right)\right]_{p=0}\phi(y)\notag\\&\hspace{2cm}\times e^{-ip\cdot(x-y)}.
\end{align}
This can be written as
\begin{align}
      &[\Pi^R\odot\phi](x) \nonumber \\
     =& \int d^4 y \int \frac{d^4p}{(2 \pi)^4} \left[ \Pi^R\left(p,\frac{x+y}{2}\right)\phi(y) e^{- i \overleftarrow{\partial}_p \cdot \overrightarrow{\partial}_y}\right]_{p=0}\notag\\&\hspace{2cm}\times e^{- i p\cdot (x-y)}.
\end{align}
Performing the $p$ integration gives a delta function $\delta(x-y)$. One can then integrate by parts with respect to $y$ to arrive at
% % \begin{align}
% = \sum_{n=0}^\infty \int d^4 y \int\frac{d^4 p}{(2\pi)^4}   \Pi^R(0,(x+y)/2)\phi(y) \frac{(-i\partial_y)^n}{n!} e^{-ip\cdot(x-y)}
% % \end{align}
\begin{align}
        % &= \sum_{n=0}^\infty \int d^4 y  \partial^n_p \Pi^R(0,(x+y)/2)\phi(y)\frac{(-i\partial_y)^n}{n!} \partial^n_y \delta(x-y)  \notag\\
        %  &= \int d^4 y   \Pi^R(p,(x+y)/2)\phi(y) e^{- i \overleftarrow{\partial}_p \cdot \overrightarrow{\partial}_y} \delta(x-y)  \notag\\
          &[\Pi^R\odot\phi](x) \nonumber \\
           =& \int d^4 y \delta(x-y)   \left[ e^{i \partial_p \cdot \partial_y}  \Pi^R\left(p,\frac{x+y}{2} \right)\phi(y) \right]_{p=0}  \notag\\
            =& \left[ e^{i \partial_p \cdot \partial_y}   \Pi^R\left(p,\frac{x+y}{2}\right)\phi(y) \right]_{p=0, y=x},
        %  &= \sum_{n=0}^\i
        %  &= \sum_{n=0}^\infty \int d^4 y \int\frac{d^4 p}{(2\pi)^4}e^{-ip\cdot(x-y)}\Pi^R(p,(x+y)/2)\phi(y)\notag\\
        % =&\int d^4 y \int\frac{d^4 p}{(2\pi)^4}e^{-ip\cdot y}\Pi^R(p,x-y/2)\phi(x-y)\notag\\
        % =&\int d^4 y \int\frac{d^4 p}{(2\pi)^4}\Pi^R(p,x-y/2) e^{-i\partial_p\cdot\partial_x} e^{-ip\cdot y} \phi(x)\notag\\
        % =&\int d^4 y \int\frac{d^4 p}{(2\pi)^4}\Pi^R(p,x-y/2) e^{i\overleftarrow{\partial_p}\cdot\overrightarrow{\partial_x}} e^{-ip\cdot y} \phi(x)
    \end{align}
which is the result \eqref{convolution:homogeneous}.

\section{Normalisation of angular momentum eigenstates}\label{App:Scattering}

%We now use

%\begin{equation}
%\int_0^\mathcal{T} dt {\rm e}^{i \omega^{\prime} t} {\rm e}^{-i (\omega - m \Omega) t} = \frac{
 %{\rm e}^{ i(\omega^{\prime} - (\omega - m \Omega))T} - }
%{i(\omega^{\prime} - (\omega - m \Omega)) }
%\end{equation}

%to obtain
%\begin{align}
%\left|\int_0^\mathcal{T} dt {\rm e}^{i \omega ^{\prime} t} {\rm e}^{-i (\omega - m \Omega) t} \right|^2 & = 4 \left(\frac{
%\sin^2\left[ \left(\omega' - (\omega - m \Omega) \right)\mathcal{T}/2\right]
%}{(\omega^{\prime}- (\omega - m \Omega))^2} \right)  \\ & \xrightarrow{\mathcal{T} \rightarrow \infty} 2 \pi \mathcal{T} \delta(\omega^{\prime} - (\omega - m \Omega)). \nonumber
%\end{align}

In this appendix we will derive the normalisation of angular momentum eigenstates given in equation \eqref{eq:denominator}. As usual in quantum field theory, delta functions arising from the normalisation of states correspond to the infinite volume of spacetime. Our free angular momentum eigenstates are normalised as
\begin{equation}
\braket{\omega \ell m \left| \omega \ell m \right. } = 2 \pi \lim_{\omega^{\prime} \rightarrow \omega} \delta(\omega - \omega^{\prime}) = \frac{2 \pi}{ v} \lim_{k^{\prime} \rightarrow k} \delta(k - k^{\prime})
\end{equation}
This arises from the normalisation spherical bessel functions associated to spherical waves:
\begin{equation}
  \int_0^\infty d r \, r^2 \,   j_\ell ( k r)  j_\ell( k' r) = \frac{\pi}{2 k^2} \delta(k - k'),
\end{equation}
which can be regulated using a finite integration range up to radius $\mathcal{R}$
\begin{align}
 & \int_0^R d r \, r^2 \,   j_\ell ( k r)  j_\ell( k r) \\&=   \frac{\pi  \mathcal{R}^2 \left(J_{l+\frac{1}{2}}(k \mathcal{R}){}^2-J_{\ell-\frac{1}{2}}(k \mathcal{R}) J_{\ell+\frac{3}{2}}(k \mathcal{R})\right)}{4 k}.  \nonumber
\end{align}
Hence we can regulate
% \begin{align}
%  \delta(k - k') 
% %  & 
% %  =  \frac{2 k^2}{\pi}\int_0^\infty d r \, r^2 \,   j_\ell ( k r)  j_\ell( k' r) \\&
%  = \lim_{R \rightarrow \infty} \frac{2 k^2}{\pi}\int_0^R d r \, r^2 \,   j_\ell ( k r)  j_\ell( k' r). \nonumber
% \end{align}
% Taking the coincidence limit of the momenta gives
\begin{align}
    & \lim_{k^{\prime} \rightarrow k} \delta(k - k^{\prime})   \nonumber \\
    % = \lim_{R \rightarrow \infty} \frac{2 k^2}{\pi}\int_0^R d r \, r^2 \,   j_\ell ( k r)  j_\ell( k r) \\& 
   & = \lim_{R \rightarrow \infty} \frac{2 k^2}{\pi}\frac{\pi  \mathcal{R}^2 \left(J_{l+\frac{1}{2}}(k \mathcal{R}){}^2-J_{l-\frac{1}{2}}(k \mathcal{R}) J_{l+\frac{3}{2}}(k R)\right)}{4 k}. 
\end{align}
We now use the asymptotic form of the Bessel function
\begin{equation}
J_{\alpha}(z) \rightarrow \sqrt{\frac{2}{\pi z}} {\rm cos} \left(z - \frac{\alpha \pi}{2} - \frac{\pi}{4} \right),
\end{equation}
for large $z$ and half-integer $\alpha$. This gives
\begin{align}
&\delta(k - k^{\prime}) \nonumber  \\
&= \lim_{\mathcal{R} \rightarrow \infty} \frac{2 k^2}{\pi} \frac{\mathcal{R}}{2 k^2} \left( {\rm sin}^2\left(k \mathcal{R} - \frac{\ell \pi}{2}\right)+{\rm cos}^2\left(k \mathcal{R} - \frac{\ell \pi}{2}\right) \right) \nonumber \\
&= \lim_{\mathcal{R} \rightarrow \infty} \frac{\mathcal{R}}{\pi} 
\end{align}

Note that for a spherical wave with energy density $\rho \propto 1/r^2$, the number of particles per unit radius (i.e. number of particles in a spherical shell of unit width) is constant. We can therefore interpret the factor $1/\mathcal{R}$ as the number of particles per unit radius. This is the spherical equivalent of the usual replacement   $(2 \pi)^3 \delta^{(3)}({\bf k}=0) \rightarrow V$.

\section{Superradiant bound states}\label{App:BoundStates}

In this appendix we detail the spherical integrals used to derive equation \eqref{eq:SRrate} in section \ref{sec:BoundStates}. Equation \eqref{eq:PabsBoundState} gives explicitly
\begin{align}
    P_{\rm abs} \simeq \int^{\mathcal T}_0 & dt dt^{\prime}  \braket{O_{J_1...J_\ell}(t^{\prime}) O_{L_1...L_\ell}(t)} \nonumber \\
    &\bra{n,l,m} \partial^{I_1}...\partial^{I_\ell} \phi(t^{\prime})\ket{0} \nonumber \\&  \nonumber \bra{0} \partial^{K_1}...\partial^{K_\ell} \phi(t) \ket{n,l,m} \\& R_{I_1}^{J_\ell}(t^{\prime})...R_{I_\ell}^{J_\ell}(t^{\prime}) R_{K_1}^{L_\ell}(t)...R_{K_l}^{L_\ell}(t).
\end{align}
 We identify the Wightman correlation function:
 \begin{equation}
     \braket{O_{J_1...J_\ell}(t^{\prime}) O_{L_1...L_\ell}(t)} = \delta_{J_1...J_\ell}^{L_1...L_\ell} \int \frac{d \omega}{2 \pi} \Delta_\ell(\omega) {\rm e}^{i \omega (t - t^{\prime})}.
 \end{equation}
Inserting this into $P_{\rm abs}$ gives:
\begin{align}
   & P_{\rm abs} = \int  \frac{d \omega^{\prime}}{2 \pi} \Delta_\ell(\omega^{\prime}) \\
   & \left| \int^{\mathcal T}_0 dt {\rm e}^{i \omega^{\prime} t} \bra{0} \partial^{K_1}...\partial^{K_\ell} \phi(t) \ket{n\ell m} R^{L_1}_{K_1}(t)...R^{L_l}_{K_\ell}(t) \right|^2, \nonumber
\end{align}
where we use a compact notation meaning that the free $K_i$ indices become contracted with themselves upon taking the modulus squared.  Using the mode expansion~Eq.~\eqref{eq:ModeExpansion}, we can write explicitly
\begin{align}
    & \bra{0} \partial^{K_1}...\partial^{K_\ell} \phi(t) \ket{n \ell m} =  \sum_{n^{\prime} l^{\prime} m^{\prime}} \frac{1}{\sqrt{2 \omega_{n^{\prime} l^{\prime}}}}\nonumber \\
   & \times \bra{0} \hat{a}_{n^{\prime} \ell^{\prime} m^{\prime}} \partial^{K_1}...\partial^{K_l} f_{n^{\prime} l^{\prime} m^{\prime}}(0) {\rm e}^{-i \omega_{n^{\prime} \ell^{\prime}}} \ket{n\ell m}. \nonumber
\end{align}
$f_{n \ell m}$ is evaluated at the spatial location of the star. Using the orthogonality of the states $\ket{ n \ell m}$, we have:
\begin{equation}
\bra{0} \partial^{K_1}...\partial^{K_\ell} \phi(t) \ket{ n \ell m} = \frac{1}{\sqrt{2 \omega_{n \ell}}} \partial^{K_1}...\partial^{K_\ell} f_{n \ell m}(0) {\rm e}^{-i \omega_{n \ell} t}.
\end{equation}
This gives
\begin{align}
   & P_{\rm abs} = \int \frac{d \omega}{2 \pi} \frac{\Delta_\ell(\omega)}{2 \omega_{\ell  n}} \nonumber  \\
   & \times \left| \int^{\mathcal T}_0 dt {\rm e}^{i (\omega - \omega_{n \ell}) t}  \partial^{I_1}...\partial^{I_\ell} f_{n \ell m}(r=0) R^{J_1}_{I_1}(t)...R^{J_\ell}_{I_\ell}(t) \right|^2. 
\end{align}
We will now use the identities \cite{Thorne:1980ru}
\begin{equation}\label{eq:Ylm}
    Y_{lm}(\theta, \phi) = \sqrt{\frac{(2\ell +1)!!}{4 \pi \ell!}} V^m_{I_1...I_\ell} \hat{r}^{I_1}...\hat{r}^{I_\ell},
\end{equation}
where
%see eqs. (1.6b) and (2.11)-(2.12) of Thorn.
\begin{align}
   & V_{I_{1} \cdots I_{\ell}}^{m}=\sum_{j=0}^{[(\ell-m) / 2]} \nu^{\ell m j}\left(\delta_{\left(I_{1}\right.}^{1}+i \delta_{\left(I_{1}\right.}^{2}\right) \cdots\Big(\delta_{I_{m}}^{1}+i \delta_{I_{m}}^{2}\Big) \nonumber \\& 
   \times \delta_{I_{m+1}}^{3} \cdots \delta_{I_{\ell-2 j}}^{3}\left(\delta_{I_{\ell-2 j+1}}^{a_{1}} \delta_{I_{\ell-2 j+2}}^{a_{1}}\right) \cdots\left(\delta_{I_{\ell-1}}^{a_{j}} \delta_{\left.I_{\ell}\right)}^{a_{j}}\right),
\end{align}
with
\begin{equation}
    \nu^{\ell m j}=\sqrt{\frac{\ell !(\ell-m) !}{(2 \ell-1) ! !(\ell+m) !}} \frac{(-1)^{m+j}(2 \ell-2 j) !}{2^{\ell} j !(\ell-j) !(\ell-m-2 j) !}.
\end{equation}
% Therefore: \textcolor{purple}{JMcD ``therefore" well the following equaiton does not ``therefore" follow, you don't use any of the previous results in deriving it...plus (see next comment) I'm not sure it makes sense to randomly put in a single derivative term on $f$. }
% \begin{align}
%     &\partial^I f_{n \ell m}(r=0) \nonumber \\
%     &= \left( [\partial^{I} R_{n \ell}(r)] Y_{lm}(\theta \phi) + R_{n \ell}(r) \partial^{I} Y_{lm} (\theta, \phi) \right) |_{r=0}.
% \end{align}
% \textcolor{purple}{JMcD - I put in the argument for r into the second $R_{n l}$. But as I say, I don't think this equation should be there.}
In order to progress, we must evaluate the multi-derivative operator acting on $f_{n l m}$ and extarct its leading order contribution. For this we shall require an expression for the bound states \cite{Day:2019bbh}
\begin{align}
&R_{n \ell} = \nonumber \\
&
\mu^{3/2} \alpha_{\ell n}^{3/2}
\sqrt{\frac{n!}{2 (n+ 2 \ell + 1)! (n + \ell +1)}}  e^{-x/2} x^{\ell} L_{n}^{2 \ell + 1} 
\left[ x \right],  \label{eq:Rnl}
\end{align}
where
\begin{align}
&x = r \mu \alpha_{\ell n}, \nonumber \\
&\alpha_{\ell n} = \frac{\mu r_s}{\ell + n  +1}, \quad \omega^2_{\ell n}  = \mu^2 \left( 1 - \frac{\alpha^2_{\ell n}}{4}\right) \simeq \mu^2  \label{eq:BSFreqs} \\
\nonumber 
\end{align}
and $L_n^{2 \ell + 1}(x)$ is a generalised Laguerre polynomial. Note that these satisfy the correct normalisation $\int dr r^2 R_{n \ell}^2 =1$. From this we can combine Eqs.~\eqref{eq:Ylm} and \eqref{eq:Rnl} to derive the leading order result in $\alpha_{\ell n}$ of the form
\begin{align}
    &\partial^{I_1}...\partial^{I_\ell} f_{n \ell m}(r=0) \simeq  \nonumber \\
    &    \sqrt{\frac{l! (2\ell +1)!!}{4 \pi }}  V^{I_1...I_\ell}_m  \left. \frac{R_{n \ell}(r)}{r^\ell} \right|_{r=0} , 
\end{align}
where we used 
\begin{align}
V^m_{J_1...J_\ell} \partial^{I_1}...\partial^{I_\ell}  r^{J_1}...r^{J_\ell} &  = l! V^m_{I_1...I_\ell}   .
\end{align}
Next, making use of the identities
\begin{align}
&V_{I_{1} \cdots I_{\ell}}^{m} R^{I_{1}}{ }_{J_{1}} \cdots R^{I_{\ell}}{ }_{J_{\ell}} =V_{J_{1} \cdots J_{\ell}}^{m} e^{i m \Omega t}, \\
\nonumber \\
&( V^m_{J_1\cdots J_\ell})^*  V_{m'}^{J_1\cdots J_\ell}  = \delta^m_{m'}
\end{align}
which leads to
\begin{align}
     P_{\rm abs} =& \frac{l! (2\ell +1)!! }{4 \pi} 
    % V^{J_1...J_\ell}_m V^{\star m}_{J_1...J_\ell} 
    \left|\frac{R_{n \ell}(r)}{r^\ell} \right|^2_{r=0} \\&
    \times\int \frac{d \omega}{2 \pi} \frac{\Delta_\ell(\omega)}{2 \omega_{\ell  n}} \left[ \int^{\mathcal T}_0 dt {\rm e}^{i(m \Omega + \omega - \mu)t} \right]^2. \nonumber
\end{align}
Explicitly we have
\begin{align}
   \left. \frac{R_{n \ell}(r)}{r^\ell} \right|_{r=0}  =&  \mu^{\ell + 3/2} \alpha_{\ell n}^{3/2 + \ell}  \sqrt{\frac{n!}{2 (n+ 2 \ell + 1)! (n + \ell +1)}}  \nonumber \\
 & \times   \frac{(2 \ell + n +1)!}{(1 + 2 \ell)! \, n! }
\end{align}
where we used $ L_{n}^{2 \ell + 1}[0]  = (1 + 2 l + n)!/[(1 + 2 l)! \, n!]$. Putting this together, for sufficiently long times we can therefore write
% \begin{align}
%     &P_{\rm abs} = \nonumber \\
%     &A_{n \ell m} \left( \frac{1}{r_{n \ell}} \right)^{2\ell +3} \int \frac{d \omega}{2 \pi} \frac{\Delta_\ell(\omega)}{2 \mu} \mathcal{T} (2 \pi) \delta(m \Omega + \omega - \mu), \nonumber \\ \textcolor{purple}{ \rm Fran}
% \end{align}
%\begin{align}
 %   A_{n \ell m} & = \frac{l!(2\ell +1)!!}{4 \pi} \left| %\frac{R_{n \ell}(r)}{r^\ell} \right|_{r=0}^2 \\& = \left( \frac{2}{a n} %\right)^{2\ell +3} \frac{l!(2\ell +1)!!}{8 \pi n (n-l-1)! (n+l)! %(2\ell +1)!}, \nonumber
%\end{align}
\begin{align}
    &P_{\rm abs}  \nonumber \\
    &=A_{n \ell m}
    % \left( \mu \alpha_{\ell n} \right)^{2\ell +3}
    \left( \frac{1}{r_{n \ell} } \right)^{2\ell +3}
    \int \frac{d \omega}{2 \pi} \frac{\Delta_\ell(\omega)}{2 \omega_{\ell  n}} \mathcal{T} (2 \pi) \delta(m \Omega + \omega - \mu),  \nonumber \\
    & = A_{n \ell m}
    % \left( \mu \alpha_{\ell n} \right)^{2\ell +3}
    \left( \frac{1}{r_{n \ell} } \right)^{2\ell +3}
    \frac{\Delta_\ell(\omega)}{2 \omega_{\ell  n}} \mathcal{T} , 
\end{align}
% \begin{align}
%     A_{n \ell m} & = \frac{l!(2\ell +1)!!(n+l)!}{8 \pi n (n-l-1)! (2\ell +1)!^2}. \textcolor{purple}{Fran}
% \end{align}
where 
\begin{equation}
r_{n\ell} = \frac{n+ l+1}{\mu^2 r_s}
\end{equation}
and
\begin{align}
    A_{n \ell m} & = \frac{\ell! (2 \ell+1)\text{!!} (2 \ell+n+1)!}{8 \pi  (\ell+n+1) n! (2 \ell+1)!^2 }.
\end{align}

% \textcolor{purple}{JMcD: note people define the principle quntum number $\bar{n} = \ell + n + 1$, which gives a critical radius $r_{c} = (\bar{n}^2/\alpha^2) r_g$ where $\alpha = r_g \mu$ with $r_g = G M$. See (1711.08298 - eq 23 -24) and (1004.3558 eq (11) ). Note however that at some point, I think people started to forget the bar and mistook $\bar{n}$ for $n$ see e.g. (2.9) of 1804.03208.}

\bibliographystyle{apsrev4-1}
\bibliography{Ref.bib}

\end{document}